\documentclass[sigconf]{acmart}
\usepackage{multirow}

\AtBeginDocument{%
  \providecommand\BibTeX{{%
    \normalfont B\kern-0.5em{\scshape i\kern-0.25em b}\kern-0.8em\TeX}}}

\setcopyright{acmcopyright}
\copyrightyear{2023}
\acmYear{2023}
\acmDOI{XXXXXXX.XXXXXXX}
\acmConference[xx]{}{xx}{xx}
\acmPrice{15.00}
\acmISBN{978-1-4503-XXXX-X/18/06}

\acmSubmissionID{5357}

\begin{document}

\title{TransRec: Learning Transferable Recommendation from Mixture-of-Modality Feedback}

\author{%
	Jie Wang$^{1,3,\ast}$
 , Fajie Yuan$^{2,\dagger}$
 ,  Mingyue Cheng$^{4}$,  Joemon M. Jose$^{1}$, 
	Chenyun Yu$^{6}$, Beibei Kong$^{3}$, Zhijin Wang$^{5}$, Bo Hu$^{3}$ , Zang Li$^{3}$
	\
}

\affiliation{%
  \institution{
  $^1$University of Glasgow, \ 
  $^2$Westlake University, \ 
  $^3$Platform and Content Group, Tencent, \  \\ 
    $^4$University of Science and Technology of China, \ 
    $^5$Jimei University, \ 
    $^6$Sun Yat-sen University
    }
    \city{}
  \country{}
}
\email{j.wang.9@research.gla.ac.uk, 
	yuanfajie@westlake.edu.cn,  
	mycheng@mail.ustc.edu.cn,}
\email{ joemon.jose@glasgow.ac.uk, 
    yuchy35@mail.sysu.edu.cn,  
	}
 \email{zhijinecnu@gmail.com,   
    {echokong, harryyfhu, gavinzli}@tencent.com
    }

\renewcommand{\shortauthors}{Trovato and Tobin, et al.}

\begin{abstract}
\renewcommand{\thefootnote}{}
    \footnotetext{$^{\ast}$Work was done when Jie Wang was a visiting scholar at Westlake University and intern at Platform and Content Group, Tencent. $^{\dagger}$Corresponding author. Fajie Yuan designed the research and Jie Wang led the experiments.}
    Learning large-scale pre-trained models on broad-ranging data and then transfer to a wide range of target tasks has become the de facto paradigm in many machine learning (ML) communities.
    Such big models are not only strong performers in practice but also offer a promising way to break out of the task-specific modeling restrictions, thereby enabling task-agnostic and unified ML systems. 
   However, such a popular paradigm is mainly unexplored by the recommender systems (RS) community. A critical issue is that standard recommendation models are primarily built on categorical identity features. That is, the users and the interacted items are represented by their unique IDs, which are generally not shareable across different systems or platforms. To pursue the transferable recommendations, we propose studying pre-trained RS models in a novel scenario where a user's interaction feedback involves a mixture-of-modality (MoM) items, e.g., text and images. 
 We then present TransRec, a very \emph{simple} modification made on the popular ID-based RS framework.
 TransRec  learns directly from the raw features of the MoM items in an end-to-end training manner and thus enables effective transfer learning
  under various scenarios without relying on overlapped users or items.
  We empirically study the transferring ability of TransRec across four different real-world recommendation settings. Besides, we look at its effects by scaling source and target data size. Our results suggest that learning neural recommendation models from MoM feedback provides a promising way to realize universal RS. 
   We will release our codes and  pre-trained
parameters for reproducibility.$^1$\footnote{$^1$We will make available the private dataset used in this study to the wider research community, however, by email, given the potential privacy and copyright issues.} 
\end{abstract}




\maketitle
\section{Introduction}
The mainstream recommender systems (RS) typically model domain-specific user behaviors and then generate item recommendations only for the same platform. Such specialized RS has been well-established in literature~\cite{cheng2016wide, covington2016deep,he2017neural,hidasi2015session,kang2018self,yuan2019simple}, yet they routinely suffer from some intrinsic limitations, such as low-accuracy problems for the cold-start setting~\cite{yuan2020parameter}, heavy manual work and high cost for training from scratch per-task models ~\cite{yuan2021one}. Hence, developing general-purpose recommendation models to be useful to many systems has significant practical value. These kinds of models are  popular in computer vision (CV)~\cite{he2016deep,dosovitskiy2020image} and natural language processing (NLP) ~\cite{devlin2018bert,brown2020language} literature, and they are recently referred to as the foundation models (FM)~\cite{bommasani2021opportunities}.Despite their remarkable progress, there has yet to be a recognized learning paradigm for building \underline{g}eneral-\underline{p}urpose  models for \underline{r}ecommender \underline{s}ystems (gpRS). 

One principal reason is that ID-based collaborative filtering (CF) techniques nearly dominate existing RS models. Thereby, well-trained RS models can only be used to serve the current system because neither users nor items are easily shared across different private systems, e.g., from TikTok\footnote{https://www.tiktok.com/} to YouTube\footnote{https://www.youtube.com/}. Even in some special cases, where userIDs or itemIDs in two platforms can be shared, it is still less prone to realizing the desired transferability since users/items on the two platforms also have limited overlapping situations. Less user and item overlapping will lead to limited transferring effects. Recent attempts such as PeterRec~\cite{yuan2020parameter}, lifelong Conure model~\cite{yuan2021one} and STAR~\cite{sheng2021one} fall exactly into this category. 

To study the transferability of RS, we attempt to explore item modality\footnote{modality used in this paper mainly refers to some multimedia modalities, such as text, image, audio or videos,  excluding the categorical IDs.}- or content-based recommendation where a modality encoder represents items (e.g. BERT~\cite{devlin2018bert} and ResNet~\cite{he2016deep}) rather than the ID embedding. By modeling modality features intuitively, recommendation models have the potential to achieve domain transferring for this modality in a broader sense --- i.e. no longer relying on overlapping and shared-ID information. Moreover, the latest revolution of robust encoder networks in NLP and CV is also potentially beneficial to modality-based item recommendation and might even bring about a paradigm shift for RS from ID-based CF back to content-based recommendation. 

To this end, we study a common yet unexplored recommendation scenario where user behaviors are composed of items with mixed-modality features --- e.g., these interacted items by a user can be all texts or images or both. However, for simplicity, each item is restricted to only one modality for this study. Such a scenario is prevalent in many practical recommender systems such as feeds recommendation, where recommended feeds can be a piece of news, an image, or a  micro-video. In this paper, we claim that \textbf{developing recommendation models based on user feedback with mixture-of-modality (MoM) items is a vital way towards transferable and general-purpose recommendation}.  To verify our claim, we design TransRec, a recommendation framework for modeling user MoM feedback. In essence,  \textbf{TransRec is model-agnostic and can be a direct modification on most ID-based recommendation models}.  To eliminate the non-transferable  ID features, we encode an item by a modality encoder rather than the itemID embedding. We encode users by a sequence of items (again, each item here is also encoded by their respective MoM encoder) rather than the userID embedding, as shown in Figure~\ref{figure2}. We evaluate TransRec on the simple yet most widely adopted two-tower-based DSSM model~\cite{huang2013learning}, where one tower represents users, and the other represents items.  We train TransRec, including both user and item encoders, by an end-to-end manner rather than using frozen features pre-extracted from modality encoders.


More importantly, we perform extensive empirical studies on TransRec --- the first RS regime enabling effective transfer across modalities \& domains. 
Specifically, we first train TransRec on a large-scale source dataset collected from a commercial website, where a user's feedback contains either textual or visual modality or both. Then we evaluate the pre-trained TransRec on the first target dataset collected from a different platform but has similar user feedback formats. Second, we evaluate TransRec on the second target dataset, where items have only one modality. Third, we evaluate TransRec with still one modality, but along with additional user/item features to verify its flexibility. At last, we evaluate TransRec on another target dataset where items look very different from the source domain to demonstrate its generality. Beyond this, we empirically examine the performance of TransRec with varying scaling strategies on both the source and target datasets. Our results confirm that TransRec effectively learns from MoM feedback for various transfer learning tasks and benefits from large-scale source and small-scale target datasets. To some extent, TransRec is by far probably the closest model toward the goal of gpRS. We believe its broad transferability will point out a new way toward the foundation models in the RS domain.

To summarize, the contributions of our work can be described as: (1) we identify an essential fact that learning from MoM feedback has the potential to realize the goal of gpRS. 
To the best of our knowledge, we are the few works of learning a universal recommendation model from over one modality feedback.
(2) we present TransRec, the first recommendation model that realizes both cross-modality and cross-domain recommendation. Note that we do not emphasize the novelty of network design since the idea of TransRec can be applied to various ID-based recommendation networks; 
(3) we  study the transferability of TransRec across four types of recommendation scenarios; (4) we  study the effects of TransRec by scaling the data, and provide some valuable insights; 
(5) we make our codes,
pre-trained parameters and training datasets of TransRec are available for future research. We hope our work can facilitate the study of applying MoM feedback to the gpRS problem. 

\begin{table}[htbp]
\setlength{\tabcolsep}{6pt}
  \centering
  \footnotesize
  \caption{Comparison of the transferability of existing recommendation methods. `O' in the source domain indicates the modality type used for pre-training and in the target domain indicates that the pre-trained model  can be used to serve recommendations with items of this modality, otherwise denoted by  `X'. 
  }
    \begin{tabular}{lccccccc}
    \toprule
    \multirow{2}[4]{*}{Methods} & \multicolumn{3}{c}{Source domain} &   & \multicolumn{3}{c}{Target domain} \\
\cmidrule{2-4}\cmidrule{6-8}          & Text  & Image  & Mixed &  & Text  & Image  & 
Mixed \\
  \midrule
    PeterRec ~\cite{yuan2020parameter} & X    & X    & X   &  & X    &  X    &  X \\
    \midrule
    ZESRec ~\cite{ding2021zero} & O    & X    & X   &  & O    &  X    &  X \\
    \midrule
    UnisRec ~\cite{hou2022towards} & O     &  X     &  X  &   & O     &  X     &  X \\
    \midrule
    CLUE ~\cite{shin2021scaling} & O     &  X     &  X  &    &  O     &  X     &  X \\
    \midrule
    TransRec (Ours) & O     & O    & O  &   & O     & O     & O \\
    \bottomrule
    \end{tabular}%
  \label{tab:addlabel}%
\end{table}%

\section{Related Work}
In this section, we briefly review the progress of gpRS and its
core techniques from two categories: self-supervised pre-training (SSP) and transfer learning (TF).

\textbf{General-purpose Recommendation.}  
Big foundation models have achieved astounding feats in the NLP and CV communities~\cite{bommasani2021opportunities}. BERT~\cite{devlin2018bert}, GPT-3~\cite{brown2020language},  ResNet~\cite{he2016deep} and various image   Transformers (ViT)~\cite{dosovitskiy2020image,arnab2021vivit, liu2021swin} have almost dominated the two fields because of their superb performance and transferability. By contrast, very few efforts have been devoted to foundation recommendation models, which are key to gpRS. 
Some work adopts multi-task learning (MTL) to combine multiple objectives to obtain a more general representation model~\cite{ma2018modeling,zhao2019recommending,tang2020progressive,ni2018perceive}. However, typical MTL is only useful to these trained tasks and cannot be directly transferred to new recommendation tasks or scenarios. 

PeterRec~\cite{yuan2020parameter} proposed the first pre-training-then-fine-tuning paradigm for learning and transferring general-purpose user representation, closely following the BERT model in NLP. Following it, Conure~\cite{yuan2021one}
introduced the `one person, one model, one world' idea and claimed that recommendation models benefit from lifelong learning. 
Similar cross-domain recommendation work also include~\cite{liu2020exploiting, ma2019pi, wu2020ptum, zhu2020deep,chen2021user,man2017cross,sheng2021one,cheng2021learning}.
However, these models are all based on the shared-ID assumption in both the source and target domains,   which, as mentioned above, tends to be too strict to hold in practice unless they are made for systems or platforms owned by the same company.

\begin{figure*}[h] 
  \centering
  \includegraphics[scale=1]{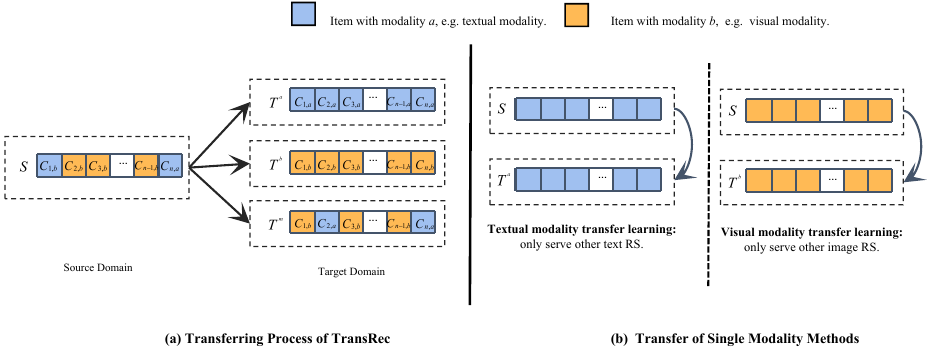}
  \caption{Schematic of learning from MoM. TransRec first pre-trains a unified recommendation model with MoM feedback in the source domain and then serves any target domain as long as the item's modality type is contained in the MoM feedback.}
  \label{figure1}
\end{figure*}

Distinct from the above work, some very recent paper (many of them are still
 preprints)  ~\cite{shin2021one4all,shin2021scaling,ding2021zero,wu2021userbert,hou2022towards,liu2022collaborative} started to devise general-purpose gpRS by leveraging the textual information, as  summarized in Table~\ref{tab:addlabel}. 
 ZESRec~\cite{ding2021zero}, CLUE~\cite{shin2021scaling} and UnisRec~\cite{hou2022towards} took a significant step forward and showed that RS models trained on textual item sequence could be effectively transferred to generate recommendations in other text recommendation scenarios. 
~\cite{wu2021mm,wu2021empowering, mu2022id} explored recommendation with text or image (with offline features extracted from ResNet~\cite{he2016deep} rather than the end-to-end training manner in TransRec) but  did not investigate the transferability issue.
The most recent preprint paper P5~\cite{geng2022recommendation} formulates various recommendation-related tasks (e.g. rating prediction, item recommendation, and explanation generation)  as a unified text-to-text paradigm with the same language modeling objective for pre-training. 
\textbf{
However, to the best of our knowledge, there exists no
gpRS models learning from multimodal or MoM feedback,\footnote{Note that learning from MoM proposed in this paper is distinct from the standard multimodal or multimedia recommendation. In our MoM scenario, an interacted item can be represented by only one single modality; but all items interacted with by an individual user could have either one or multiple modalities; by contrast, in the typical multimodal recommendation scenario, an item should contain at least two modalities simultaneously, and all items interacted by a user usually contain the same type of multimodal features as the individual item.} along with an end-to-end training fashion. 
}

\textbf{Transfer Learning.} 
Recently, there has been a growing interest in transfer learning  with
an upstream self-supervised pre-training (SSP) and downstream supervised finetuning paradigm. In terms of SSP, existing work can be broadly categorized  into two classes: BERT/GPT-like generative pre-training~\cite{devlin2018bert,brown2020language} and CPC/SimCLR-like contrastive pre-training~\cite{van2018representation,chen2020simple}. Compared with supervised learning,  SSP methods are more likely to learn general-purpose representation since they are directly trained by self-generated labels rather than those task-specific labels. Meanwhile, recommender systems with large amounts of implicit user feedback are a natural fit for self-supervised learning. Both generative pre-training and contrastive pre-training are popular in RS literature; however, much of them ~\cite{zhou2020s3,sun2019bert4rec}  only investigated pre-training for the current task than pursuing general-purpose recommendations. 
Regarding downstream fine-tuning, current work mainly adopted  parameter freezing~\cite{he2016vbpr},  full finetuning~\cite{devlin2018bert,he2016ups}, adapter  finetuning~\cite{yuan2020parameter,houlsby2019parameter} and prompt~\cite{brown2020language,gao2020making} to adapt 
an upstream SSP model to downstream tasks. This paper combines generative and contrastive pre-training strategies for learning the gpRS model and applying full finetuning for domain adaption.

\begin{figure*}
  \centering
  \includegraphics[width=1\textwidth]{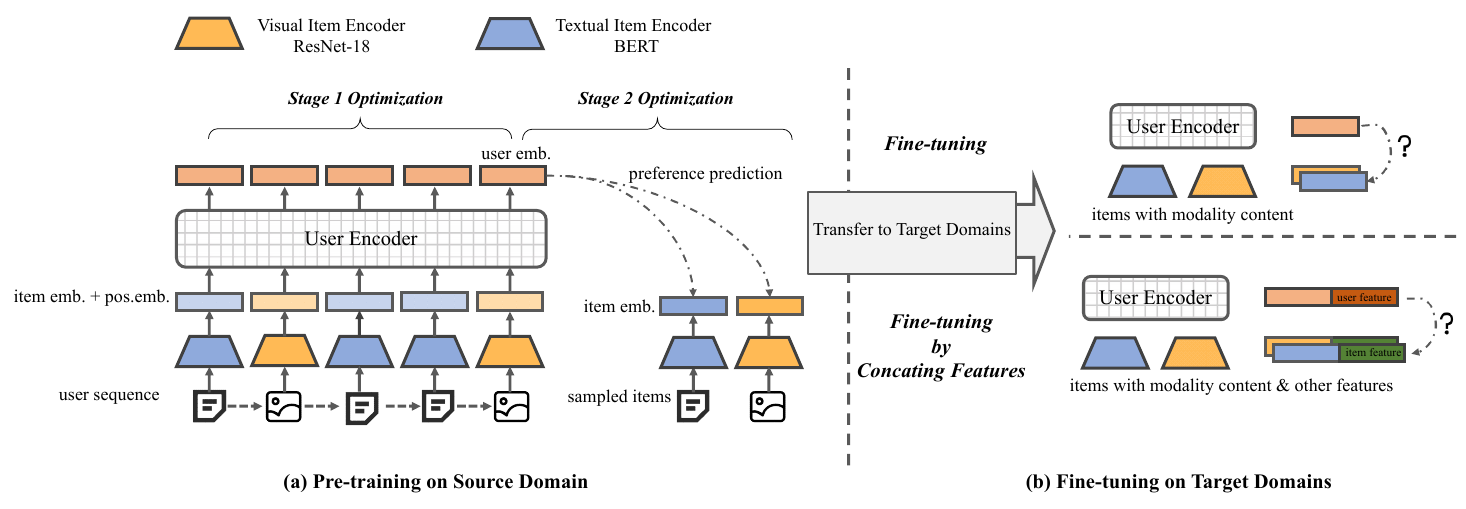}
  \caption{Illustration of the training process of TransRec. 
  Here, the inner product is employed to compute the preference between users and candidate items. }
  \label{figure2}
\end{figure*}

\section{The TransRec Framework}
In this section, we first formulate the recommendation tasks with mixture-modality feedback and introduce some notations used in this work. Then, we introduce the TransRec framework in detail.


\label{gen_inst}


\subsection{Problem Definition}


Assume that we are given two categories of domains: source domain $S$ and target domains $T=\{T_1, T_2, ..., T_N\}$. In source (target) domain, suppose that there exist user set $\mathop { U}_s$ ($\mathop {U}_t $) and item set $\mathop { V}_s$ ($\mathop { V}_t$), involving $\left| {\mathop {U}_s } \right|$ ($\left| {\mathop { U}_t } \right|$) users and $\left| {\mathop {V}_s } \right|$ ($\left| {\mathop { V}_t } \right|$) items.
In both domains, the content feature of items is recorded with modality set $\mathcal{M}=\{a, b\}$, containing textual and visual modalities, denoted as $a$ and $b$, respectively. Following the setting of item-based collaborative filtering~\cite{yuan2020parameter}, users can be represented with the sequence of their historical interaction records $\mathop {C_u} = \{c_{1,m}, ..., c_{n,m}\}$. Here, $n$ indicates the sequence length while $m\in \mathcal{M}$. This work aims to learn a generic recommender from source domain $S$ which can be transferred to $N$ target domains $T$, involving non-overlapping of userIDs or itemIDs.

As illustrated in Figure~\ref{figure1}, by learning from  ${ M}$,
the trained model can be applied to the following domains, including single-modality domain $\mathop C  = \{ \mathop c_{1,a} ,...,\mathop c_{n,a} \}$ ($\mathop c_{i,a}  \in \mathop {V}_t^a$) in domain $\mathop T^a $,
single-modality domain $\mathop C  = \{ \mathop c _{1,b} ,...,\mathop c _{n,b} \}$ ($\mathop c _{i,b}  \in \mathop { V}_t^b$) in domain $\mathop T^b$,
and mixed-modality domain $\mathop C  = \{ \mathop c_{1,m} ,...,\mathop c_{n,m} \}$ ( $\mathop c_{i,m}  \in \mathop { V}_t^m$, $m \in { M}$) in domain $\mathop T^m$.
Suppose that we extend the source domain with four modalities ${ M} = \left\{ {a,b, c,d} \right\}$ $=$ $\rm \left\{ {text, image, audio, video} \right\}$. The target domain can be served with 15 types of modalities, i.e. $\left\{ {a} \right\}$, $\left\{ {b} \right\}$, $\left\{ {c} \right\}$, $\left\{ {a,b} \right\}$, $\left\{ {a,b,c} \right\}$...$\left\{ {a,b,c,d} \right\}$ that covers a majority of existing multimedia modalities.  In other words, learning from MoM feedback could easily cover downstream recommendation scenarios with items in the form of any single or combined modalities used in the source domain, which differs from traditional recommendations where user feedback contains modality features exactly the same as the per-interacted item.



\subsection{TransRec Architecture}
\label{section_3.2}
To verify our claim, we develop TransRec on the most popular two-tower-based recommendation architecture,  a.k.a. DSSM~\cite{huang2013learning}. To eliminate ID features, we represent both users and items with item modality contents. That is, the item tower of DSSM is represented by an item modality encoder (e.g. BERT for textual items and ResNet for visual items), and the user tower is represented by a user encoder that
directly models an ordered collection of item interactions.
For this study, we choose the self-attention-based Transformer blocks as the user encoder, given their popularity and outstanding performance in modeling sequence data. But we emphasize that TransRec is not limited to the basic Transformer block --- i.e. any deep neural network module that can 
model ID embedding sequences, such as the pure MLP network~\cite{li2022mlp4rec},
GNN~\cite{chang2021sequential}, RNN~\cite{hidasi2015session}, TCN~\cite{yuan2019simple} and  more  advanced Transformers~\cite{lan2019albert,beltagy2020longformer} could be employed here (see Figure~\ref{figure2} for the TransRec architecture).
Formally, given a user interaction sequence $ \boldsymbol{\mathop C} $ from an MoM scenario, TransRec divides it into two sub-sequence $\boldsymbol{\mathop C\nolimits^u}$ and $\boldsymbol{\mathop C\nolimits^e}$, their relationship is denoted as:
\setlength{\abovedisplayskip}{10pt}
\begin{align} 
\label{splitsequence}
\boldsymbol{\mathop C } =\boldsymbol{\mathop C\nolimits^u} \cup \boldsymbol{\mathop C\nolimits^e }
\end{align}
\setlength{\belowdisplayskip}{10pt}

\noindent where $ \boldsymbol{\mathop C} = \{ \mathop c\nolimits_{1,m} ,...,\mathop c\nolimits_{n,m},\mathop c\nolimits_{n + 1,m}, ..., \mathop c\nolimits_{n + l,m} \}$, with length of $n+l$, which is divided into $\boldsymbol{\mathop C\nolimits^u}  = \{ \mathop c\nolimits_{1,m} ,...,\mathop c\nolimits_{n,m} \}$, with the front ordered $n$ interactions, and $\boldsymbol{\mathop C\nolimits^e}  = \{ \mathop c\nolimits_{n + 1,m} ,...,\mathop c\nolimits_{n + l,m} \}$, with last $l$ interactions. TransRec takes them as inputs to the item encoder $\mathop E\nolimits_i$. Then, we obtain 
$\boldsymbol{\mathop Z\nolimits^u}$
and $\boldsymbol{\mathop Z\nolimits^e}$ 
, which are item representations for 
 $\boldsymbol{\mathop C\nolimits^u}$ and $\boldsymbol{\mathop C\nolimits^e}$, respectively.
The item representations in $\boldsymbol{\mathop Z\nolimits^u}$  are then fed into the user encoder $\mathop E\nolimits_u$ to achieve user representation $\boldsymbol{\mathop U\nolimits^u}$.
$\boldsymbol{\mathop U\nolimits^u}$ and $\boldsymbol{\mathop Z\nolimits^e}$ are used to compute their relevance scores $\boldsymbol{\mathop R\nolimits_{u,e}}$. The process can be formulated as: 
\begin{align} 
\label{process}
&\boldsymbol {\mathop Z\nolimits^e}  = \mathop E\nolimits_i \left(\boldsymbol{{\mathop C\nolimits^e } }\right), \ \ 
\boldsymbol{\mathop Z\nolimits^u}  = \mathop E\nolimits_i \left( \boldsymbol{{\mathop C\nolimits^u } }\right), \\
\label{process2}
&\boldsymbol{\mathop U\nolimits^u } = \mathop E\nolimits_u \left( \boldsymbol{{\mathop Z\nolimits^u } }\right), \ \ 
\boldsymbol{\mathop R\nolimits_{u,e}}  = \boldsymbol{\mathop U\nolimits^u } \cdot \boldsymbol{\mathop Z\nolimits^e},
\end{align}
where $\boldsymbol{\mathop Z\nolimits^u} {\rm{ = \{ }}\mathop z\nolimits_1 ,...,\mathop z\nolimits_n {\rm{\} }}$ and $\boldsymbol{\mathop Z\nolimits^e} {\rm{ = \{ }}\mathop z\nolimits_{n + 1} ,...,\mathop z\nolimits_{n + l} {\rm{\} }}$, 
$\mathop z\nolimits_{t}$ is the item representation of \textit{t}-th item in the user sequence $\boldsymbol{\mathop C}$. 
And $\boldsymbol{\mathop R\nolimits_{u,e}}  = \left\{ {\mathop r\nolimits_{u,1} ,\mathop r\nolimits_{u,2} ,\mathop {...,r}\nolimits_{u,l} } \right\}$,  
$\mathop r\nolimits_{u,t}$ denotes the relevance score between $\boldsymbol{\mathop U\nolimits^u}$ and \textit{t}-th item of the sub-sequence $\boldsymbol{\mathop C\nolimits^e}$. $\boldsymbol{\mathop R\nolimits_{u,e}}$  demonstrates the relation between the user and his next interaction sequence. Next, we describe the two components of our model in detail. 

\textbf{Item Encoder.}    
Given the MoM scenario, each item encoder of TransRec takes the item's (single) modality as input. We consider at most two types of modalities, i.e. textual tokens and image pixels, for each user in this paper.
Attempts for more modality scenarios are interesting for future work. For an item with textual modality (i.e. $\boldsymbol{\mathop c\nolimits_{i,a}}$ ), we  adopt BERT-base \cite{devlin2018bert} to model its token content $ \boldsymbol{ \mathop c\nolimits_{i,a}} = \left[ {\mathop t\nolimits_1 ,\mathop t\nolimits_2 ,...,\mathop t\nolimits_k } \right]$. Then we apply a commonly used attention network to compute the relations of each token, similar as in ~\cite{wu2021empowering}, and average pooling is used to obtain the final textual representation $\boldsymbol{\mathop Z\nolimits_{i}^{}}$:
\begin{align}\label{eq5}
\boldsymbol{\mathop Z\nolimits_{i}^{}} = {\mathop{\rm Avg}\nolimits}({\mathop{\rm SelfAtt}\nolimits}({\mathop{\rm BERT}\nolimits} (\boldsymbol{ \mathop c\nolimits_{i,a}}))).
\end{align}

\noindent Likewise, we apply the ResNet-18 \cite{he2016deep} to encoder an item with visual modality shown by an image $\boldsymbol{\mathop c\nolimits_{i,b}}$. Then we perform average pooling to produce the final visual representation $\boldsymbol{\mathop Z\nolimits_{i}^{} }$, followed by a standard MLP layer, which is given:
\begin{align}\label{eq6}
\boldsymbol{\mathop Z\nolimits_{i}^{} } = {\mathop{\rm Avg}\nolimits}({\mathop{\rm MLP}\nolimits} ({\mathop{\rm ResNet}\nolimits} (\boldsymbol{\mathop c\nolimits_{i,b}} ))).
\end{align}
 
\textbf{User Encoder.}
For user encoder, we propose using the Transformer  (denoted as $\mathop{\rm Trsf_{u}}$) block architecture, where each token embedding is the representation from an item encoder rather than the original word ID embedding. Position embedding $\boldsymbol{P} = \{ \mathop p\nolimits_1 ,...,\mathop p\nolimits_n \}$ is added to model the sequential  patterns of user behaviors.
The specific process is formulated as follows:
\begin{align}
\label{user}
&\boldsymbol{\mathop S\nolimits^u } = \boldsymbol{\mathop Z\nolimits^u } + \boldsymbol{\mathop P\nolimits^u}, \\
\label{userbert}
&\boldsymbol{\mathop U\nolimits^u}  = \mathop E\nolimits_u^{} \left( {\boldsymbol{\mathop S\nolimits^u }} \right) =  {\mathop{\rm Last}\nolimits}({\mathop{\rm Trsf_{u}}\nolimits}(\boldsymbol{\mathop S\nolimits^u} )).
\end{align}
where the user representation $\mathop U\nolimits^u$ is represented by the hidden state at the last position of the user encoder. 

\subsection{Optimization}
\label{optimization}
Inspired by the pre-training and fine-tuning paradigm, we apply a similar training regime for TransRec: first pre-training the user encoder network, and then training the whole framework of TransRec.

\textbf{Stage 1: User Encoder Pre-training.} 
We perform pre-training for user encoder network in a self-supervised manner. 
Specifically, we apply the left-to-right style generative pre-training to predict the next item in the interaction sequence, similarly as SASRec and NextItNet~\cite{kang2018self,yuan2019simple}. The way we choose unidirectional pre-training rather than BERT-style (bidirectional) (i.e. Eq.(\ref{userbert})) is simply because unidirectional pre-training converges much faster but without obvious loss of precision. We use the softmax cross-entropy loss as the objective function:
\begin{align}
\label{s1}
&\boldsymbol{\mathop {\tilde y}\nolimits_t}{\rm{ = }}{\mathop{\rm softmax}\nolimits} ({\mathop{\rm RELU}\nolimits} (\boldsymbol{\mathop {S'}\nolimits_t} \boldsymbol{\mathop W\nolimits^U}  + \boldsymbol{\mathop b\nolimits^U} )),\\
&\boldsymbol{\mathop { L}\nolimits_{{\rm{UEP}}}}  =  - \sum\limits_{u \in U} {\sum\limits_{t \in \left[ {1,...,n} \right]}^{} {\left( {\boldsymbol{\mathop y\nolimits_t^{} } \log \left( {\boldsymbol{\mathop {\tilde y}\nolimits_t }} \right)} \right)} },
\end{align}
where $\boldsymbol{\mathop W^U}$, $\boldsymbol{\mathop b^U}$ are the projection matrix \& bias terms, $\boldsymbol{\mathop {S'}_t}$ is the representation of  last hidden layer.

\textbf{Stage 2: End-to-End Training.}  
TransRec is trained in an end-to-end manner by fine-tuning model parameters of both user and item encoders.
This is vastly different from much  multimodal recommendation literature where they first pre-extract offline features by modality encoder and then treat them as fixed features for a recommendation network~\textcolor{black}{\cite{he2016vbpr, wei2019mmgcn, yi2022multi}}. End-to-end training enables a better adaption of textual and visual features to the current recommendation domain. 
Specifically, we
propose to use the Contrastive Predictive Coding (CPC) ~\cite{van2018representation} learning method. 
Given a sequence of user interactions $\boldsymbol{\mathop C} $, we divide the sequence into sub-sequence $\boldsymbol{\mathop C\nolimits^u}$ and $\boldsymbol{\mathop C\nolimits^e}$
to encode the relationship between them.
The binary cross entropy loss function  is as follows:
\begin{align}
\label{eq11}
\boldsymbol{\mathop { L}\nolimits_{{\rm{CPC}}}}  =  - \sum\limits_{u \in U} {\left[ {\sum\limits_{t = n + 1}^{n + l} {\log \left( {\sigma \left( {\boldsymbol{\mathop r\nolimits_{u,t}} } \right)} \right) + \sum\limits_{g = 1}^{j} {\log \left( {1 - \sigma \left( {\boldsymbol{\mathop r\nolimits_{u,g} }} \right)} \right)} } } \right]}
\end{align}
where $\boldsymbol{\mathop {_g}}$ is randomly  sampled negative items~\cite{rendle2012bpr,yuan2016lambdafm} during model training. For each user sequence, we choose 4 un-interactive items from the whole recommendation pool as negatives.

\subsection{Transfer to Downstream Tasks}


After the training process of TransRec, the user \& item representations and their matching relationship in the source domain can be  well learned, which  can then be used to improve various downstream recommendation tasks by performing a direct fine-tuning on the corresponding task.

The fine-tuning process of TransRec in the target domain follows the same end-to-end training procedure (i.e. Stage 2 of Section~\ref{optimization}) as used in the source domain. The key difference is that (i) TransRec could achieve faster and better convergence by fine-tuning on the target domain because of its well pre-trained parameters; (ii) TransRec could use fewer training examples to achieve the same performance compared to itself without pre-training.
Because of the flexible architecture of DSSM, TransRec can also incorporate additional user and item features like many standard ID-based models. That is, we can fuse user/item features by simply concatenating them with user/item representation generated by the user/item encoder network (also see Figure~\ref{figure2} (b)). 

During the online inference stage, TransRec is as efficient as ID-based models since the modality representations of all items can be calculated in advance by an offline manner.

\section{Empirical Study}
To verify the effectiveness of our proposed TransRec, we conduct empirical experiments and evaluate pre-trained recommenders on four types of downstream tasks. 
\label{headings}
\subsection{Experiments for The Source Domain}
\label{imp}
\textbf{Datasets.}
The source data is the news recommendation data collected from QQBrowser\footnote{https://browser.qq.com/} from $14$th to $17$th, December $2020$. We collect around 25 million user-item interaction behaviors, involving about 1 million randomly sampled users and 133, 000 interacted items. Each interaction denotes observed feedback at a certain time, including full play and clicks. The users with less than 20 interactions are removed and the average length of user sequences is 25.


\begin{table}[] \footnotesize
\setlength{\tabcolsep}{10.2pt}
\caption{Characteristics of the source dataset. 
`Form' indicates the item category. `All' means all users in this dataset, including the above three types.
For example, the first line denotes that there are 765,895 users whose interacted items always have two-modal (i.e. textual and visual) features.
}
\label{table1}
\centering
\begin{tabular}{lcccc}
\toprule
Form    & Modality          & User               & Item             & Interaction        \\ \midrule
Mixed   & Text + Image         & 765,895            & 133,107                & 19,233,882          \\
Article & Text              & 133,107            & 62,837                & 3,327,463           \\
Video   & Image              & 123,897            & 70,270              & 2,996,048           \\ \midrule
All     & Text + Image         & 1,022,899          & 133,107          & 25,557,393 \\ \bottomrule
\end{tabular}
\end{table}

We construct the sequence behaviors for each user using his recent $32$ ordered interactions. 
Beyond the ID features, our datasets 
represent each item with its raw modality  features. 
More accurately,  items in a user session can be videos-only,  news-only, or both. However, each item is either a video or a piece of news, i.e. containing only one modality.
We adopt item titles to represent news, item thumbnails to
represent videos.
 The statistics are in Table~\ref{table1}.

\textbf{Evaluation Metrics.}
We use the typical \textit{leave-one-out} strategy \cite{he2017neural} for evaluation, where the last item of the user interaction sequence is denoted as test data, and the item before the last one is used as validation data. 
The remaining sequence is used as the training data, in which the latest five interactions are used to predict given all previous interactions (see Figure~\ref{figure2}).
We pad zero at the end of the user sub-sequences $\boldsymbol{\mathop C\nolimits^u}$ if their lengths are smaller than 25. Unlike many previous methods that employ a small scope of randomly sampled items for evaluation, which may lead to inconsistencies with the non-sampled version \cite{krichene2020sampled}, we rank the full item set without using the inaccurate sampling measures.
We apply Hit Ratio (HR)~\cite{yuan2020parameter} and Normalized Discounted Cumulative Gain (NDCG)~\cite{yuan2019simple} to measure the performance of each method. Our evaluation methods are consistent for both the source and downstream tasks.

\textbf{Implement Details.}
 Hyper-parameters are searched on the validation set to ensure a fair evaluation. We notice that the Adam~\cite{kingma2014adam} optimizer performs the best on all validation sets. Then we set the learning rate to $1e^{-4}$, embedding size to $256$ and batch size to  $512$ after empirical searching and training TransRec on 4 NVIDIA A100s (40G memory).  
We train all models until it converges and saves parameters when they reach the highest accuracy on the validation set. We set the layer number of the user encoder to $4$ and the head number of multi-head attention to $4$. The input size of an image is 3×224×224, and the token length of an article description is set to 32 maximum.

\textbf{Results.} 
Before evaluating TransRec on the downstream datasets, we first examine its performance in the source domain. 
The purpose is not to demonstrate that TransRec can achieve state-of-the-art recommendation accuracy. Since it depends on both user and item encoder networks --- i.e. with a more expressive user and item encoders, TransRec is likely to be more powerful. Instead, we hope to investigate via fair comparison,  whether learning from modality contents has advantages over traditional ID-based methods i.e. IDRec. Throughout this paper, we use IDRec to denote the counterpart of TransRec with a 
similar recommendation architecture with the same sampling and optimization method but instead with an ID embedding item encoder.


\begin{table}[] \footnotesize
\setlength{\tabcolsep}{7pt}
\caption{Results on the source dataset. The terms below have the same meaning as in Table~\ref{table1}.
TransRec- denotes TransRec without first stage user encoder pre-training. 
}
\label{table2}
\centering
\begin{tabular}{lccccc}
\toprule
Method                      & Modality    & HR@5     & NDCG@5    & HR@10    & NDCG@10  \\ \midrule
IDRec                       & ID        & 0.0141    & 0.0089    & 0.0230    & 0.0118    \\ \midrule
\multirow{4}{*}{TransRec-}  & Image      & 0.0337    & 0.0216    & 0.0540	& 0.0281 \\
                            & Text      & 0.0330	& 0.0214    & 0.0536	& 0.0280 \\
                            & Mixed     & 0.0326    & 0.0210    & 0.0530	& 0.0275\\
                            & All       & 0.0327    & 0.0211   & 0.0532    & 0.0276   \\ \midrule
\multirow{4}{*}{TransRec}   & Image     & 0.0696     & 0.0414   & 0.1128    & 0.0553 \\
                            & Text     & 0.0325    & 0.0206    & 0.0582    & 0.0272    \\
                            & Mixed    & 0.0390    &0.0233	  & 0.0679    & 0.0326   \\
                            & All  	   & 0.0403   & 0.0239     & 0.0699      & 0.0334    \\ \bottomrule
                    
\end{tabular}
\end{table}

Table \ref{table2} shows the results of IDRec and TransRec regarding \textcolor{black}{HR@\{5, 10\} and NDCG@\{5, 10\}}. Please note `TransRec-' denotes TransRec without the first stage user encoder pre-training. Two important observations can be made: (1)  TransRec- and TransRec largely outperform IDRec (e.g. 0.0699 vs. 0.0230, and 0.0532 vs. 0.0230 on HR@10), demonstrating strong potential of learning from modality content data. The higher results of TransRec are presumably attributed to three key factors:  large-scale training data, powerful item encoder and user encoder networks, and an end-to-end training fashion.
(2) TransRec exceeds TransRec- with all types of modality settings (e.g. 0.0699 vs. 0.0532, and 0.0679 vs. 0.0530 on HR@10), which evidences the effectiveness of user encoder pre-training (see Section~\ref{optimization}). 
The results of (1) further motivate us to develop modality content-based recommendation models for downstream tasks.
\textcolor{black}{\textbf{Training Comparison.}
In Table \ref{table_flp}, we present TransRec \emph{vs.} IDRec in terms of FLOPs (the number of floating-point operations).
As can be seen, TransRec requires almost 2000$\times$ larger computing cost than IDRec alongside up to 5$\times$ more training time. TransRec, with the standard training method, achieves better accuracy than IDRec, but with the cost of much higher computing and training time.  
This result is not surprising since pre-training foundation models in other fields, such as BERT-large, GPT-3, and various ViT require huge computing power. We believe the results here will inspire more research to study model optimization and compression techniques for large gpRS --- e.g. fine-tuning only the top few layers or applying more advanced fine-tuning techniques, such as adapter~\cite{yuan2020parameter,houlsby2019parameter} and prompt tuning~\cite{lester2021power}.} 


\begin{table} \footnotesize
\setlength{\tabcolsep}{6pt}
\caption{Comparison of TransRec (full parameter fine-tuning) and IDRec in terms of FLOPs and training time per batch. TransRec is trained on a user sequence with 13 images and 12 news articles where images have a size of 3$\times$224$\times$224 and text length is 32.}
\label{table_flp}
\centering
\begin{tabular}{lccc}
\toprule
Model        & FLOPs        & s/batch                          \\ \midrule
IDRec      & 32.973M    & 0.7s                               \\
TransRec   & 78.305G  &  3.2s                           \\ \bottomrule
\end{tabular}
\end{table}

\label{finetune}

\subsection{Experiments for The Target Domains}
All the target datasets below are from other recommender systems.

\textbf{TN-mixed:} 
It was collected from Tencent News (TN)\footnote{https://news.qq.com/}, where an interacted item can be either a News article or a video  thumbnail. Similar to the source domain, the interacted item set of a user contains mixture-of-modality features, i.e. both text and images.

\begin{table} \footnotesize
\setlength{\tabcolsep}{11pt}
\caption{Datasets for downstream recommendation tasks.  Except for DouYin, interactions in other datasets are mainly about users' clicking or watching behaviours.
}
\label{table3}
\centering
\begin{tabular}{lcccc}
\toprule
Domain        & Modality    & User    & Item    & Interaction                       \\ \midrule
TN-mixed      & Text+Image   & 49,639  & 48,383  & 870,894                                \\
TN-video      & Image      & 47,004  & 50,053   & 809,483                           \\
TN-text       & Text        & 49,033  & 49,142 & 903,543 \\
DouYin        & Image        & 100,000 & 66,228 &  1,688,944                              \\ \bottomrule
\end{tabular}
\end{table}

\textbf{TN-video and TN-text:} The two datasets only contain items with a single modality. For example,
users' interactions in TN-video include only videos, while  TN-text includes only textual items, i.e. news. They are used to evaluate TransRec's generality for single-modal item recommendation.
Given that users and items in real-world recommender systems have various additional features, 
we introduce two types of user features (gender and age) and one item feature (category) for TN-text.
\begin{table*}[]
\caption{Comparison of recommendation results on four downstream domains.  
TFS denotes training target datasets with random parameters as initialization. It shares the same network architecture and hyper-parameters as TransRec.
The best results are bolded.}
\label{table4}
\centering
\setlength{\tabcolsep}{8pt}
\footnotesize
\begin{tabular}{l|c|ccccccc}
\toprule

Domain                      & Modality                       & Metric & IDRNN   & IDCNN       & IDRec  & TFS
& TransRec  &Improv. \\ \midrule
\multirow{4}{*}{TN-mixed}   & \multirow{4}{*}{Mixed}   & HR@5  & 0.0109   & 0.0112      & 0.0117    & \underline{0.0249} & \textbf{0.0285} &14.46\%\\                                 
                                     &                         & NDCG@5 & 0.0063   &0.0067   & 0.0068       & \underline{0.0160} & \textbf{0.0177} &10.63\%\\
                                     &                         & HR@10  & 0.0129  & 0.0195   & 0.0210       & \underline{0.0428} & \textbf{0.0478} &11.68\%\\
                                     &                         & NDCG@10 & 0.0062  & 0.0094   & 0.0100     & \underline{0.0213} & \textbf{0.0239} &12.21\%\\ \midrule
\multirow{4}{*}{TN-video}   & \multirow{4}{*}{Image  }         & HR@5 & 0.0134   & 0.0159   & 0.0153       & \underline{0.0208}      & \textbf{0.0271} &30.29\%\\                                 
                                     &                         & NDCG@5  & 0.0093 & 0.0098  & 0.0092     & \underline{0.0131}       & \textbf{0.0173} &30.06\%\\
                                     &                         & HR@10  & 0.0201 & 0.0265    & 0.0267    & \underline{0.0336}      & \textbf{0.0424} &26.19\%\\
                                     &                        & NDCG@10 & 0.0114 & 0.0133   & 0.0125    &\underline{0.0173}      & \textbf{0.0221} &27.75\%\\ \midrule
                                 
\multirow{4}{*}{TN-text}   & \multirow{4}{*}{Text}          & HR@5  & 0.0105   & 0.0123  & 0.0105       & \underline{0.0303} & \textbf{0.0358} & 18.15\%\\                                 
                                      &                      & NDCG@5   & 0.0063  & 0.0078    & 0.0062    & \underline{0.0192} & \textbf{0.0227} &18.23\%\\
                                      &                      & HR@10    & 0.0189  & 0.0220   & 0.0192    & \underline{0.0500} & \textbf{0.0597} &19.40\%\\
                                      &                      & NDCG@10  & 0.0089  & 0.0109  & 0.0090    & \underline{0.0255} & \textbf{0.0303} &18.82\%\\ \midrule
\multirow{4}{*}{DouYin}    & \multirow{4}{*}{Image  }      & HR@5    &0.0059    &0.0057   &0.0023  & \underline{0.0115} & \textbf{0.0146}&26.96\% \\                               
                                      &                  & NDCG@5   &0.0037    & 0.0035    &0.0014    & \underline{0.0073} & \textbf{0.0090} &23.29\%\\
                                      &                  & HR@10    & 0.0096   & 0.0100    &0.0035     & \underline{0.0205} & \textbf{0.0259} &26.34\%\\
                                      &                  & NDCG@10  &  0.0049  & 0.0049   & 0.0018    & \underline{0.0101} & \textbf{0.0126} &24.75\%\\ \bottomrule
\end{tabular}
\end{table*}

\begin{figure}[h]
  \centering
  \includegraphics[scale=0.3]{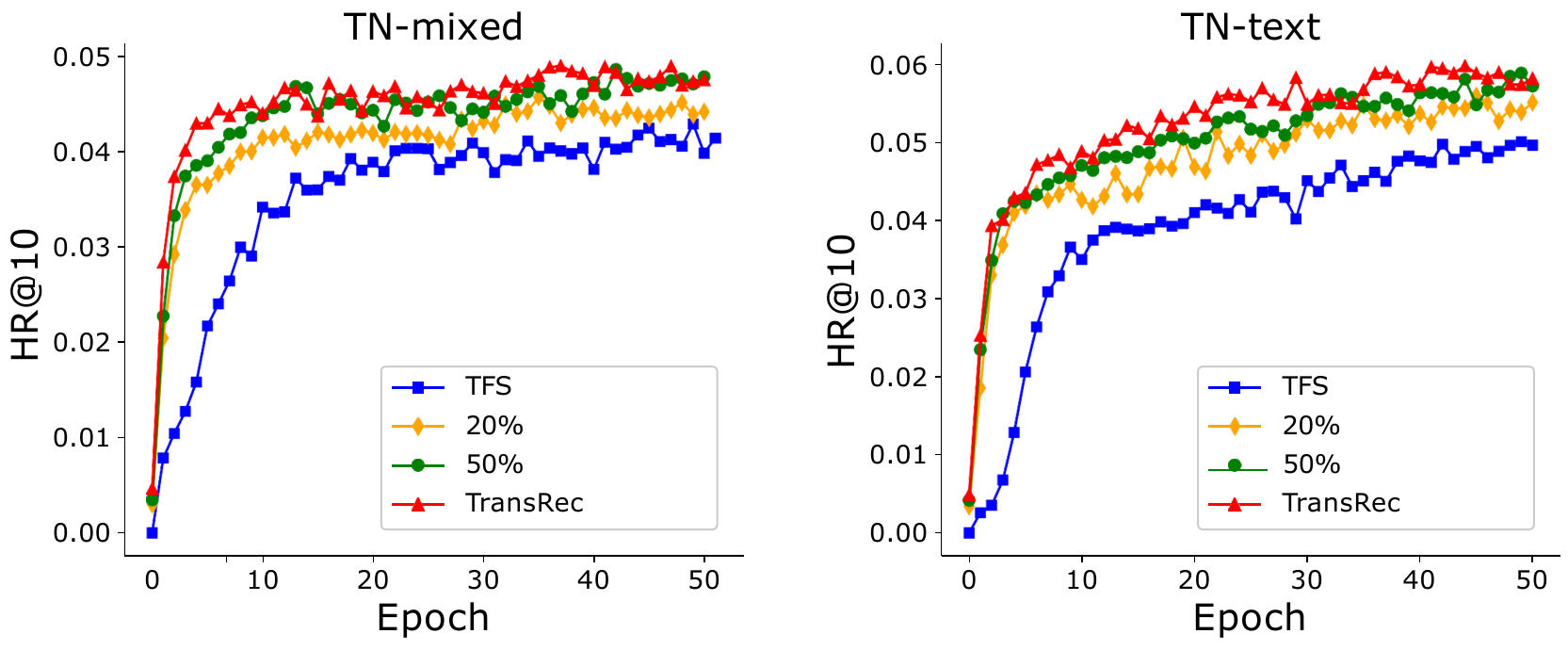}
  \caption{Convergence trend by scaling the source data.}
  \label{figure3}
\end{figure}

\textbf{DouYin:} It was collected from DouYin\footnote{https://www.douyin.com/} (the Chinese version of TikTok), a well-known short video recommendation application. Unlike all previous datasets, the positive user feedback in DouYin only contains comment behaviors. In addition, video genres and cover image size in DouYin are vastly different from the source domain.
Table \ref{table3} summarizes the statistics of downstream  datasets.

\textbf{Baselines.} 
The key baseline used to compare with TransRec is TFS, which is used to verify whether TransRec pre-trained in the source domain can effectively improve the performance of the downstream tasks. Thereby, TFS adopts the same architecture, hyper-parameters and modality inputs as TransRec, but is trained from scratch with randomly initialized parameters for the key architecture (i.e. the user encoder). 
Note that its item encoder still adopts the pre-trained BERT and ResNet. In addition, we compare TransRec with  several ID-based baselines for reference, including IDRNN~\cite{hidasi2015session}, IDCNN~\cite{yuan2019simple} and IDRec.
IDRNN uses the GRU to encode user sequence, while IDCNN and IDRec use the temporal CNN (TCN) and Transformer architecture to encode user sequence. 

For a fair comparison, we adopt IDRNN, IDCNN and IDRec with exactly the same training objective function, negative sampling method and optimization strategy as the fine-tuned TransRec (see Figure~\ref{figure2}). \textcolor{black}{All the batch size of target domains is set to 128 and the learning rate of TN-mixed, TN-video, TN-text and Douyin is 
set to $1e^{-5}$, $1e^{-5}$, $1e^{-5}$, and $5e^{-5}$, respectively.}
Despite that, we emphasize again that the purpose of this study is neither to propose a more advanced neural recommendation architecture nor to pursue some state-of-the-art results.  
The key purpose of this study is to indicate that: (1) learning from modality content features instead of ID features achieves the goal of transferable recommendations across
different domains; (2) learning from MoM feedback rather than single-modal or typical multimodal feedback reaches the goal of generic recommendations across different modalities.

\textbf{Results.}
The overall results are shown in Table \ref{table4}, which includes four recommendation scenarios.
Two important observations can be made: (1) TransRec performs consistently better than its training-from-scratch version, i.e. TFS; 
(2) TransRec performs better than ID-based methods as well. 
The results suggest that training the source domain brings much better results for TransRec on all target datasets. By analyzing these scenarios, we can conclude that TransRec --- learning from MoM feedback --- can be broadly transferred to various recommendation scenarios, including the source-like
mixed-modal scenario (i.e. TN-mixed), the single-modal scenario (TN-video), the scenario with more additional features (TN-text), and the scenario with very different modality content (DouYin). 

\subsection{Scaling Effects}
\label{scaling}


\textbf{Scaling effects of the source dataset.}  Table \ref{table6} shows the model performance in the four downstream recommendation tasks. First, it can be seen that the recommendation accuracy of TransRec is improved by scaling up the source training data. For example,  HR@10 in the TN-mixed dataset grows from 0.0428 to 0.0448 with 20\% of the source data, and then it grows from 0.0448 to 0.0474 with 50\% of the source data. 
 Such a property of TransRec is desired since it implies that scaling up the 
 source dataset is an effective way to improve downstream tasks. We also plot the convergence behavior in Figure \ref{figure3}, which shows consistent improvements.


\begin{table}[]
\caption{Results of TransRec by scaling up the source corpus. 
}
\label{table6} \footnotesize
\setlength{\tabcolsep}{8pt}
\centering
\begin{tabular}{l|ccccc}
\toprule
Domain                         & Metric  & TFS & 20\%   & 50\%   & TransRec  \\ \midrule
\multirow{2}{*}{TN-mixed}        & HR@10   & 0.0428       & 0.0448 & 0.0474 & 0.0478 \\
                                 & NDCG@10 & 0.0213       & 0.0227 & 0.0237 & 0.0239 \\ \midrule
\multirow{2}{*}{TN-video}        & HR@10   & 0.0336       & 0.0400 & 0.0417 & 0.0424 \\
                                 & NDCG@10 & 0.0173       & 0.0209 & 0.0214 & 0.0221 \\ \midrule
\multirow{2}{*}{TN-text}         & HR@10   & 0.0500       & 0.0543 & 0.0581 & 0.0597 \\
                                 & NDCG@10 & 0.0255       & 0.0281 & 0.0292 & 0.0303 \\ \midrule
\multirow{2}{*}{DouYin}          & HR@10   & 0.0205       & 0.0233 & 0.0254 & 0.0259 \\
                                 & NDCG@10 & 0.0101       & 0.0113 & 0.0120 & 0.0126 \\ \bottomrule
\end{tabular}
\end{table}

\begin{figure*}[t]
  \centering
  \includegraphics[width=0.85\textwidth]{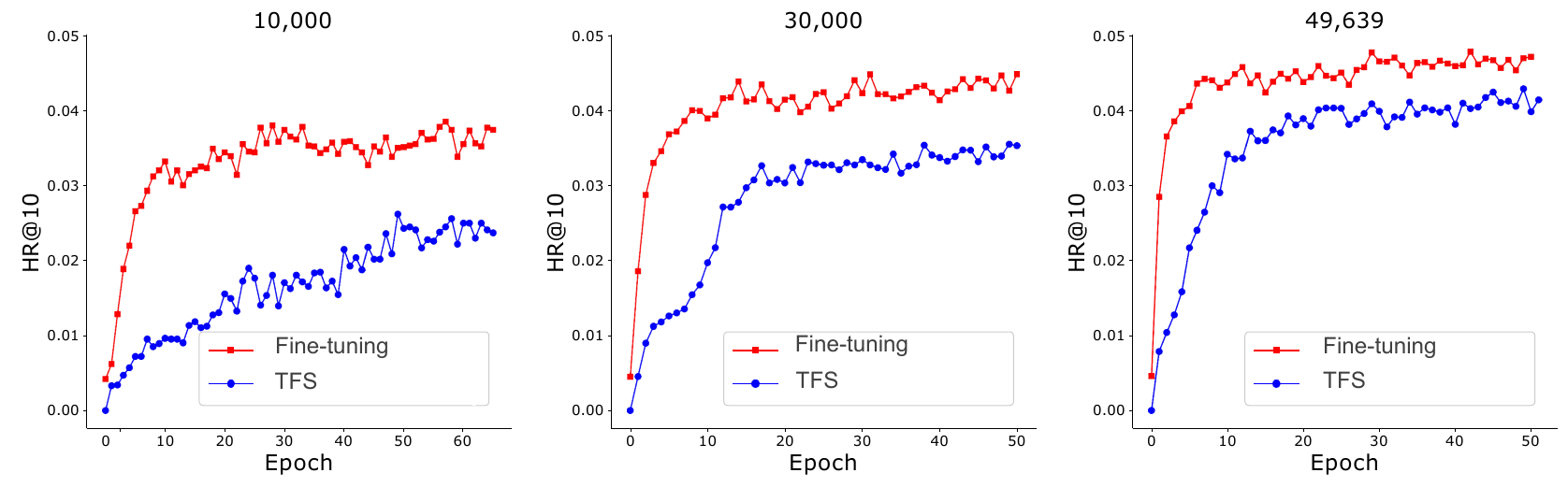}
  \caption{Comparison of convergence by scaling TN-mixed dataset.}
  \label{figure4}
\end{figure*}

\textbf{Scaling effects of the target dataset.} 
We study the effects of TransRec by scaling down the target data, aiming to verify whether TransRec can alleviate the insufficient data issue. Specifically, we decrease the size of these target datasets by using 20\% and 60\% of the original data. 
Results are shown in Table~\ref{table7} and Figure~\ref{figure4}. It can be seen that (1) TransRec's accuracy increases with more training data; (2) more accuracy gains are achieved  with less
training data. This suggests that when a recommender system lacks training data, transferring the (user-item) matching relationship from a large source dataset is helpful. 


\begin{table}[] \footnotesize
 \setlength{\tabcolsep}{4pt}
\caption{Comparison of relative performance improvement on downstream tasks with varied target data size. `Num. Sample' denotes the number of user behavior sequences for training.
`Improv.' indicates the relative performance improvement of TransRec compared with TFS.  }

\label{table7}
\centering
\begin{tabular}{@{}lccccccc@{}}
\toprule
\multirow{2}{*}{Domain}     & \multirow{2}{*}{\begin{tabular}[c]{@{}c@{}}Num. \\ Sample\end{tabular}} & \multicolumn{2}{c}{HR@10} & \multirow{2}{*}{Improv.} & \multicolumn{2}{c}{NDCG@10} & \multirow{2}{*}{Improv.} \\ \cmidrule(lr){3-4} \cmidrule(lr){6-7}
                          &                                                                         & {TFS}   & TransRec   &                          & {TFS}    & TransRec    &                          \\ \midrule
\multirow{3}{*}{TN-mixed} & 10,000                                                                  & 0.0261        & 0.0385    & 41.51\%                  & 0.0126         & 0.0193     & 53.17\%                  \\
                          & 30,000                                                                  & 0.0354        & 0.0448    & 26.55\%                  & 0.0176         & 0.0223     & 26.70\%                  \\
                          & 49,639                                                                  & 0.0428        & 0.0478    & 11.68\%                   & 0.0213         & 0.0239     & 12.21\%                  \\ \midrule
\multirow{3}{*}{TN-text}  & 10,000                                                                  & 0.0393        & 0.0597    & 51.91\%                  & 0.0201         & 0.0262     & 30.35\%                  \\
                          & 30,000                                                                  & 0.0453        & 0.0549    & 21.19\%                  & 0.0230         & 0.0283     & 23.04\%                  \\
                          & 49,033                                                                  & 0.0500        & 0.0597    & 19.40\%                  & 0.0255         & 0.0303     & 18.82\%                  \\ \bottomrule
\end{tabular}
\end{table}

\begin{table} \footnotesize
\setlength{\tabcolsep}{1.3pt}
\centering
\caption{ End-to-end training vs. frozen features.   
}
\label{table9}
\begin{tabular}{lccccccccc} 
\toprule
\multirow{2}{*}{Method}             & \multirow{2}{*}{Manner} & \multicolumn{2}{c}{TN-mixed} &  & \multicolumn{2}{c}{TN-text} &  & \multicolumn{2}{c}{TN-video}  \\ 
\cmidrule{3-4}\cmidrule{6-7}\cmidrule{9-10}
                                    &                         & HR@10  & NDCG@10             &  & HR@10  & NDCG@10            &  & HR@10  & NDCG@10              \\ 
\midrule
\multirow{2}{*}{TFS} 
                                    & Frozen                 & 0.0334 & 0.0167              &  & 0.0350  & 0.0176             &  & 0.0037 & 0.0017               \\
                                    & End2end                 & 0.0428 & 0.0213              &  & 0.0500   & 0.0255            &  & 0.0336 & 0.0173               \\ 
\midrule
\multirow{2}{*}{TransRec}           & Frozen                 & 0.0359 & 0.0181              &  & 0.0411 & 0.0206             &  & 0.0040 & 0.0023               \\
                                    & End2end                 & 0.0478 & 0.0239             &  & 0.0597 & 0.0303             &  & 0.0424 & 0.0221               \\
\bottomrule
\end{tabular}
\end{table}

\subsection{End-to-End Training v.s. Frozen Features.}
\label{sec4_4}

\textcolor{black}{Traditional multimodal and multimedia recommendations have been well studied~\cite{he2016vbpr}. While due to high computing resources and a less powerful text/image encoder network, prior art tends to extract frozen modality features and feed them into a CTR or recommendation model. Such practice is prevalent for industrial applications given billions of training examples~\cite{covington2016deep,cheng2016wide}. However, we 
want to explore whether end-to-end learning is superior to learning from frozen features. The results are in Table~\ref{table9}.
We can achieve consistent improvements through end-to-end learning; however,  finetuning  BERT and ResNet is more computationally expensive than using pre-extracted features. Notably, we notice that frozen textual features yield worse results than visual features. This may imply that
the textual features generated by BERT are more general than the visual features generated by ResNet.
This finding is also aligned with findings in NLP and CV fields --- finetuning all parameters is generally better than finetuning only the classification layer (with the backbone network frozen). }

\section{Conclusion and Future Works}
In this paper, we study a novel recommendation scenario where user feedback contains items with mixture-of-modality features. We develop TransRec, the first recommendation model learning from MoM feedback in an end-to-end manner with the goal of learning general-purpose models for recommender systems. To show its transferring ability, we conduct an empirical study in four types of downstream recommendation tasks.
Our results verify that TransRec is a generic model that can be broadly transferred to improve many recommendation tasks as long as the modality has been trained in the source domain. Our work has significant practical implications towards universal recommender systems to realize `One Model to Serve All'~\cite{yuan2021one,sheng2021one}.

One limitation is that we only examine TransRec with two types of modality features (image and text). As a result, it can only serve three scenarios: image-only, text-only, and image-text. Since both video and audio data can be represented by images~\cite{li2019neural,likhosherstov2021polyvit}, intuitively, TransRec can be extended to scenarios where items involve more modalities. For example, suppose four distinct modalities (image, text, audio and video) are available in user feedback. TransRec can potentially serve at most 15 types of scenarios covering most modalities for multimedia data. 
This is an interesting future direction to realize a more general recommender system. A second limitation is the high training cost of TransRec because of the end-to-end learning paradigm. 
Although TransRec enables effective transfer learning for many downstream recommendation tasks, its huge computation cost and training time should receive more future research attention.


\medskip

{
\small
\bibliographystyle{ACM-Reference-Format}
\bibliography{ref.bib}


\begin{thebibliography}{59}


\ifx \showCODEN    \undefined \def \showCODEN     #1{\unskip}     \fi
\ifx \showDOI      \undefined \def \showDOI       #1{#1}\fi
\ifx \showISBNx    \undefined \def \showISBNx     #1{\unskip}     \fi
\ifx \showISBNxiii \undefined \def \showISBNxiii  #1{\unskip}     \fi
\ifx \showISSN     \undefined \def \showISSN      #1{\unskip}     \fi
\ifx \showLCCN     \undefined \def \showLCCN      #1{\unskip}     \fi
\ifx \shownote     \undefined \def \shownote      #1{#1}          \fi
\ifx \showarticletitle \undefined \def \showarticletitle #1{#1}   \fi
\ifx \showURL      \undefined \def \showURL       {\relax}        \fi
\providecommand\bibfield[2]{#2}
\providecommand\bibinfo[2]{#2}
\providecommand\natexlab[1]{#1}
\providecommand\showeprint[2][]{arXiv:#2}

\bibitem[Arnab et~al\mbox{.}(2021)]%
        {arnab2021vivit}
\bibfield{author}{\bibinfo{person}{Anurag Arnab}, \bibinfo{person}{Mostafa
  Dehghani}, \bibinfo{person}{Georg Heigold}, \bibinfo{person}{Chen Sun},
  \bibinfo{person}{Mario Lu{\v{c}}i{\'c}}, {and} \bibinfo{person}{Cordelia
  Schmid}.} \bibinfo{year}{2021}\natexlab{}.
\newblock \showarticletitle{Vivit: A video vision transformer}. In
  \bibinfo{booktitle}{\emph{Proceedings of the IEEE/CVF International
  Conference on Computer Vision}}. \bibinfo{pages}{6836--6846}.
\newblock


\bibitem[Beltagy et~al\mbox{.}(2020)]%
        {beltagy2020longformer}
\bibfield{author}{\bibinfo{person}{Iz Beltagy}, \bibinfo{person}{Matthew~E
  Peters}, {and} \bibinfo{person}{Arman Cohan}.}
  \bibinfo{year}{2020}\natexlab{}.
\newblock \showarticletitle{Longformer: The long-document transformer}.
\newblock \bibinfo{journal}{\emph{arXiv preprint arXiv:2004.05150}}
  (\bibinfo{year}{2020}).
\newblock


\bibitem[Bommasani et~al\mbox{.}(2021)]%
        {bommasani2021opportunities}
\bibfield{author}{\bibinfo{person}{Rishi Bommasani}, \bibinfo{person}{Drew~A
  Hudson}, \bibinfo{person}{Ehsan Adeli}, \bibinfo{person}{Russ Altman},
  \bibinfo{person}{Simran Arora}, \bibinfo{person}{Sydney von Arx},
  \bibinfo{person}{Michael~S Bernstein}, \bibinfo{person}{Jeannette Bohg},
  \bibinfo{person}{Antoine Bosselut}, \bibinfo{person}{Emma Brunskill},
  {et~al\mbox{.}}} \bibinfo{year}{2021}\natexlab{}.
\newblock \showarticletitle{On the opportunities and risks of foundation
  models}.
\newblock \bibinfo{journal}{\emph{arXiv preprint arXiv:2108.07258}}
  (\bibinfo{year}{2021}).
\newblock


\bibitem[Brown et~al\mbox{.}(2020)]%
        {brown2020language}
\bibfield{author}{\bibinfo{person}{Tom Brown}, \bibinfo{person}{Benjamin Mann},
  \bibinfo{person}{Nick Ryder}, \bibinfo{person}{Melanie Subbiah},
  \bibinfo{person}{Jared~D Kaplan}, \bibinfo{person}{Prafulla Dhariwal},
  \bibinfo{person}{Arvind Neelakantan}, \bibinfo{person}{Pranav Shyam},
  \bibinfo{person}{Girish Sastry}, \bibinfo{person}{Amanda Askell},
  {et~al\mbox{.}}} \bibinfo{year}{2020}\natexlab{}.
\newblock \showarticletitle{Language models are few-shot learners}.
\newblock \bibinfo{journal}{\emph{Advances in neural information processing
  systems}}  \bibinfo{volume}{33} (\bibinfo{year}{2020}),
  \bibinfo{pages}{1877--1901}.
\newblock


\bibitem[Chang et~al\mbox{.}(2021)]%
        {chang2021sequential}
\bibfield{author}{\bibinfo{person}{Jianxin Chang}, \bibinfo{person}{Chen Gao},
  \bibinfo{person}{Yu Zheng}, \bibinfo{person}{Yiqun Hui},
  \bibinfo{person}{Yanan Niu}, \bibinfo{person}{Yang Song},
  \bibinfo{person}{Depeng Jin}, {and} \bibinfo{person}{Yong Li}.}
  \bibinfo{year}{2021}\natexlab{}.
\newblock \showarticletitle{Sequential recommendation with graph neural
  networks}. In \bibinfo{booktitle}{\emph{Proceedings of the 44th International
  ACM SIGIR Conference on Research and Development in Information Retrieval}}.
  \bibinfo{pages}{378--387}.
\newblock


\bibitem[Chen et~al\mbox{.}(2021)]%
        {chen2021user}
\bibfield{author}{\bibinfo{person}{Lei Chen}, \bibinfo{person}{Fajie Yuan},
  \bibinfo{person}{Jiaxi Yang}, \bibinfo{person}{Xiangnan He},
  \bibinfo{person}{Chengming Li}, {and} \bibinfo{person}{Min Yang}.}
  \bibinfo{year}{2021}\natexlab{}.
\newblock \showarticletitle{User-specific Adaptive Fine-tuning for Cross-domain
  Recommendations}.
\newblock \bibinfo{journal}{\emph{IEEE Transactions on Knowledge and Data
  Engineering}} (\bibinfo{year}{2021}).
\newblock


\bibitem[Chen et~al\mbox{.}(2020)]%
        {chen2020simple}
\bibfield{author}{\bibinfo{person}{Ting Chen}, \bibinfo{person}{Simon
  Kornblith}, \bibinfo{person}{Mohammad Norouzi}, {and}
  \bibinfo{person}{Geoffrey Hinton}.} \bibinfo{year}{2020}\natexlab{}.
\newblock \showarticletitle{A simple framework for contrastive learning of
  visual representations}. In \bibinfo{booktitle}{\emph{International
  conference on machine learning}}. PMLR, \bibinfo{pages}{1597--1607}.
\newblock


\bibitem[Cheng et~al\mbox{.}(2016)]%
        {cheng2016wide}
\bibfield{author}{\bibinfo{person}{Heng-Tze Cheng}, \bibinfo{person}{Levent
  Koc}, \bibinfo{person}{Jeremiah Harmsen}, \bibinfo{person}{Tal Shaked},
  \bibinfo{person}{Tushar Chandra}, \bibinfo{person}{Hrishi Aradhye},
  \bibinfo{person}{Glen Anderson}, \bibinfo{person}{Greg Corrado},
  \bibinfo{person}{Wei Chai}, \bibinfo{person}{Mustafa Ispir}, {et~al\mbox{.}}}
  \bibinfo{year}{2016}\natexlab{}.
\newblock \showarticletitle{Wide \& deep learning for recommender systems}. In
  \bibinfo{booktitle}{\emph{Proceedings of the 1st workshop on deep learning
  for recommender systems}}. \bibinfo{pages}{7--10}.
\newblock


\bibitem[Cheng et~al\mbox{.}(2021)]%
        {cheng2021learning}
\bibfield{author}{\bibinfo{person}{Mingyue Cheng}, \bibinfo{person}{Fajie
  Yuan}, \bibinfo{person}{Qi Liu}, \bibinfo{person}{Xin Xin}, {and}
  \bibinfo{person}{Enhong Chen}.} \bibinfo{year}{2021}\natexlab{}.
\newblock \showarticletitle{Learning Transferable User Representations with
  Sequential Behaviors via Contrastive Pre-training}. In
  \bibinfo{booktitle}{\emph{2021 IEEE International Conference on Data Mining
  (ICDM)}}. IEEE, \bibinfo{pages}{51--60}.
\newblock


\bibitem[Covington et~al\mbox{.}(2016)]%
        {covington2016deep}
\bibfield{author}{\bibinfo{person}{Paul Covington}, \bibinfo{person}{Jay
  Adams}, {and} \bibinfo{person}{Emre Sargin}.}
  \bibinfo{year}{2016}\natexlab{}.
\newblock \showarticletitle{Deep neural networks for youtube recommendations}.
  In \bibinfo{booktitle}{\emph{Proceedings of the 10th ACM conference on
  recommender systems}}. \bibinfo{pages}{191--198}.
\newblock


\bibitem[Devlin et~al\mbox{.}(2018)]%
        {devlin2018bert}
\bibfield{author}{\bibinfo{person}{Jacob Devlin}, \bibinfo{person}{Ming-Wei
  Chang}, \bibinfo{person}{Kenton Lee}, {and} \bibinfo{person}{Kristina
  Toutanova}.} \bibinfo{year}{2018}\natexlab{}.
\newblock \showarticletitle{Bert: Pre-training of deep bidirectional
  transformers for language understanding}.
\newblock \bibinfo{journal}{\emph{arXiv preprint arXiv:1810.04805}}
  (\bibinfo{year}{2018}).
\newblock


\bibitem[Ding et~al\mbox{.}(2021)]%
        {ding2021zero}
\bibfield{author}{\bibinfo{person}{Hao Ding}, \bibinfo{person}{Yifei Ma},
  \bibinfo{person}{Anoop Deoras}, \bibinfo{person}{Yuyang Wang}, {and}
  \bibinfo{person}{Hao Wang}.} \bibinfo{year}{2021}\natexlab{}.
\newblock \showarticletitle{Zero-shot recommender systems}.
\newblock \bibinfo{journal}{\emph{arXiv preprint arXiv:2105.08318}}
  (\bibinfo{year}{2021}).
\newblock


\bibitem[Dosovitskiy et~al\mbox{.}(2020)]%
        {dosovitskiy2020image}
\bibfield{author}{\bibinfo{person}{Alexey Dosovitskiy}, \bibinfo{person}{Lucas
  Beyer}, \bibinfo{person}{Alexander Kolesnikov}, \bibinfo{person}{Dirk
  Weissenborn}, \bibinfo{person}{Xiaohua Zhai}, \bibinfo{person}{Thomas
  Unterthiner}, \bibinfo{person}{Mostafa Dehghani}, \bibinfo{person}{Matthias
  Minderer}, \bibinfo{person}{Georg Heigold}, \bibinfo{person}{Sylvain Gelly},
  {et~al\mbox{.}}} \bibinfo{year}{2020}\natexlab{}.
\newblock \showarticletitle{An image is worth 16x16 words: Transformers for
  image recognition at scale}.
\newblock \bibinfo{journal}{\emph{arXiv preprint arXiv:2010.11929}}
  (\bibinfo{year}{2020}).
\newblock


\bibitem[Gao et~al\mbox{.}(2020)]%
        {gao2020making}
\bibfield{author}{\bibinfo{person}{Tianyu Gao}, \bibinfo{person}{Adam Fisch},
  {and} \bibinfo{person}{Danqi Chen}.} \bibinfo{year}{2020}\natexlab{}.
\newblock \showarticletitle{Making pre-trained language models better few-shot
  learners}.
\newblock \bibinfo{journal}{\emph{arXiv preprint arXiv:2012.15723}}
  (\bibinfo{year}{2020}).
\newblock


\bibitem[Geng et~al\mbox{.}(2022)]%
        {geng2022recommendation}
\bibfield{author}{\bibinfo{person}{Shijie Geng}, \bibinfo{person}{Shuchang
  Liu}, \bibinfo{person}{Zuohui Fu}, \bibinfo{person}{Yingqiang Ge}, {and}
  \bibinfo{person}{Yongfeng Zhang}.} \bibinfo{year}{2022}\natexlab{}.
\newblock \showarticletitle{Recommendation as Language Processing (RLP): A
  Unified Pretrain, Personalized Prompt \& Predict Paradigm (P5)}.
\newblock \bibinfo{journal}{\emph{arXiv preprint arXiv:2203.13366}}
  (\bibinfo{year}{2022}).
\newblock


\bibitem[He et~al\mbox{.}(2016)]%
        {he2016deep}
\bibfield{author}{\bibinfo{person}{Kaiming He}, \bibinfo{person}{Xiangyu
  Zhang}, \bibinfo{person}{Shaoqing Ren}, {and} \bibinfo{person}{Jian Sun}.}
  \bibinfo{year}{2016}\natexlab{}.
\newblock \showarticletitle{Deep residual learning for image recognition}. In
  \bibinfo{booktitle}{\emph{Proceedings of the IEEE conference on computer
  vision and pattern recognition}}. \bibinfo{pages}{770--778}.
\newblock


\bibitem[He and McAuley(2016a)]%
        {he2016ups}
\bibfield{author}{\bibinfo{person}{Ruining He} {and} \bibinfo{person}{Julian
  McAuley}.} \bibinfo{year}{2016}\natexlab{a}.
\newblock \showarticletitle{Ups and downs: Modeling the visual evolution of
  fashion trends with one-class collaborative filtering}. In
  \bibinfo{booktitle}{\emph{proceedings of the 25th international conference on
  world wide web}}. \bibinfo{pages}{507--517}.
\newblock


\bibitem[He and McAuley(2016b)]%
        {he2016vbpr}
\bibfield{author}{\bibinfo{person}{Ruining He} {and} \bibinfo{person}{Julian
  McAuley}.} \bibinfo{year}{2016}\natexlab{b}.
\newblock \showarticletitle{VBPR: visual bayesian personalized ranking from
  implicit feedback}. In \bibinfo{booktitle}{\emph{Proceedings of the AAAI
  Conference on Artificial Intelligence}}, Vol.~\bibinfo{volume}{30}.
\newblock


\bibitem[He et~al\mbox{.}(2017)]%
        {he2017neural}
\bibfield{author}{\bibinfo{person}{Xiangnan He}, \bibinfo{person}{Lizi Liao},
  \bibinfo{person}{Hanwang Zhang}, \bibinfo{person}{Liqiang Nie},
  \bibinfo{person}{Xia Hu}, {and} \bibinfo{person}{Tat-Seng Chua}.}
  \bibinfo{year}{2017}\natexlab{}.
\newblock \showarticletitle{Neural collaborative filtering}. In
  \bibinfo{booktitle}{\emph{Proceedings of the 26th international conference on
  world wide web}}. \bibinfo{pages}{173--182}.
\newblock


\bibitem[Hidasi et~al\mbox{.}(2015)]%
        {hidasi2015session}
\bibfield{author}{\bibinfo{person}{Bal{\'a}zs Hidasi},
  \bibinfo{person}{Alexandros Karatzoglou}, \bibinfo{person}{Linas Baltrunas},
  {and} \bibinfo{person}{Domonkos Tikk}.} \bibinfo{year}{2015}\natexlab{}.
\newblock \showarticletitle{Session-based recommendations with recurrent neural
  networks}.
\newblock \bibinfo{journal}{\emph{arXiv preprint arXiv:1511.06939}}
  (\bibinfo{year}{2015}).
\newblock


\bibitem[Hou et~al\mbox{.}(2022)]%
        {hou2022towards}
\bibfield{author}{\bibinfo{person}{Yupeng Hou}, \bibinfo{person}{Shanlei Mu},
  \bibinfo{person}{Wayne~Xin Zhao}, \bibinfo{person}{Yaliang Li},
  \bibinfo{person}{Bolin Ding}, {and} \bibinfo{person}{Ji-Rong Wen}.}
  \bibinfo{year}{2022}\natexlab{}.
\newblock \showarticletitle{Towards Universal Sequence Representation Learning
  for Recommender Systems}. In \bibinfo{booktitle}{\emph{Proceedings of the
  28th ACM SIGKDD Conference on Knowledge Discovery and Data Mining}}.
  \bibinfo{pages}{585--593}.
\newblock


\bibitem[Houlsby et~al\mbox{.}(2019)]%
        {houlsby2019parameter}
\bibfield{author}{\bibinfo{person}{Neil Houlsby}, \bibinfo{person}{Andrei
  Giurgiu}, \bibinfo{person}{Stanislaw Jastrzebski}, \bibinfo{person}{Bruna
  Morrone}, \bibinfo{person}{Quentin De~Laroussilhe}, \bibinfo{person}{Andrea
  Gesmundo}, \bibinfo{person}{Mona Attariyan}, {and} \bibinfo{person}{Sylvain
  Gelly}.} \bibinfo{year}{2019}\natexlab{}.
\newblock \showarticletitle{Parameter-efficient transfer learning for NLP}. In
  \bibinfo{booktitle}{\emph{International Conference on Machine Learning}}.
  PMLR, \bibinfo{pages}{2790--2799}.
\newblock


\bibitem[Huang et~al\mbox{.}(2013)]%
        {huang2013learning}
\bibfield{author}{\bibinfo{person}{Po-Sen Huang}, \bibinfo{person}{Xiaodong
  He}, \bibinfo{person}{Jianfeng Gao}, \bibinfo{person}{Li Deng},
  \bibinfo{person}{Alex Acero}, {and} \bibinfo{person}{Larry Heck}.}
  \bibinfo{year}{2013}\natexlab{}.
\newblock \showarticletitle{Learning deep structured semantic models for web
  search using clickthrough data}. In \bibinfo{booktitle}{\emph{Proceedings of
  the 22nd ACM international conference on Information \& Knowledge
  Management}}. \bibinfo{pages}{2333--2338}.
\newblock


\bibitem[Kang and McAuley(2018)]%
        {kang2018self}
\bibfield{author}{\bibinfo{person}{Wang-Cheng Kang} {and}
  \bibinfo{person}{Julian McAuley}.} \bibinfo{year}{2018}\natexlab{}.
\newblock \showarticletitle{Self-attentive sequential recommendation}. In
  \bibinfo{booktitle}{\emph{2018 IEEE International Conference on Data Mining
  (ICDM)}}. IEEE, \bibinfo{pages}{197--206}.
\newblock


\bibitem[Kingma and Ba(2014)]%
        {kingma2014adam}
\bibfield{author}{\bibinfo{person}{Diederik~P Kingma} {and}
  \bibinfo{person}{Jimmy Ba}.} \bibinfo{year}{2014}\natexlab{}.
\newblock \showarticletitle{Adam: A method for stochastic optimization}.
\newblock \bibinfo{journal}{\emph{arXiv preprint arXiv:1412.6980}}
  (\bibinfo{year}{2014}).
\newblock


\bibitem[Krichene and Rendle(2020)]%
        {krichene2020sampled}
\bibfield{author}{\bibinfo{person}{Walid Krichene} {and}
  \bibinfo{person}{Steffen Rendle}.} \bibinfo{year}{2020}\natexlab{}.
\newblock \showarticletitle{On sampled metrics for item recommendation}. In
  \bibinfo{booktitle}{\emph{Proceedings of the 26th ACM SIGKDD international
  conference on knowledge discovery \& data mining}}.
  \bibinfo{pages}{1748--1757}.
\newblock


\bibitem[Lan et~al\mbox{.}(2019)]%
        {lan2019albert}
\bibfield{author}{\bibinfo{person}{Zhenzhong Lan}, \bibinfo{person}{Mingda
  Chen}, \bibinfo{person}{Sebastian Goodman}, \bibinfo{person}{Kevin Gimpel},
  \bibinfo{person}{Piyush Sharma}, {and} \bibinfo{person}{Radu Soricut}.}
  \bibinfo{year}{2019}\natexlab{}.
\newblock \showarticletitle{Albert: A lite bert for self-supervised learning of
  language representations}.
\newblock \bibinfo{journal}{\emph{arXiv preprint arXiv:1909.11942}}
  (\bibinfo{year}{2019}).
\newblock


\bibitem[Lester et~al\mbox{.}(2021)]%
        {lester2021power}
\bibfield{author}{\bibinfo{person}{Brian Lester}, \bibinfo{person}{Rami
  Al-Rfou}, {and} \bibinfo{person}{Noah Constant}.}
  \bibinfo{year}{2021}\natexlab{}.
\newblock \showarticletitle{The power of scale for parameter-efficient prompt
  tuning}.
\newblock \bibinfo{journal}{\emph{arXiv preprint arXiv:2104.08691}}
  (\bibinfo{year}{2021}).
\newblock


\bibitem[Li et~al\mbox{.}(2022)]%
        {li2022mlp4rec}
\bibfield{author}{\bibinfo{person}{Muyang Li}, \bibinfo{person}{Xiangyu Zhao},
  \bibinfo{person}{Chuan Lyu}, \bibinfo{person}{Minghao Zhao},
  \bibinfo{person}{Runze Wu}, {and} \bibinfo{person}{Ruocheng Guo}.}
  \bibinfo{year}{2022}\natexlab{}.
\newblock \showarticletitle{MLP4Rec: A Pure MLP Architecture for Sequential
  Recommendations}.
\newblock \bibinfo{journal}{\emph{arXiv preprint arXiv:2204.11510}}
  (\bibinfo{year}{2022}).
\newblock


\bibitem[Li et~al\mbox{.}(2019)]%
        {li2019neural}
\bibfield{author}{\bibinfo{person}{Naihan Li}, \bibinfo{person}{Shujie Liu},
  \bibinfo{person}{Yanqing Liu}, \bibinfo{person}{Sheng Zhao}, {and}
  \bibinfo{person}{Ming Liu}.} \bibinfo{year}{2019}\natexlab{}.
\newblock \showarticletitle{Neural speech synthesis with transformer network}.
  In \bibinfo{booktitle}{\emph{Proceedings of the AAAI Conference on Artificial
  Intelligence}}, Vol.~\bibinfo{volume}{33}. \bibinfo{pages}{6706--6713}.
\newblock


\bibitem[Likhosherstov et~al\mbox{.}(2021)]%
        {likhosherstov2021polyvit}
\bibfield{author}{\bibinfo{person}{Valerii Likhosherstov},
  \bibinfo{person}{Anurag Arnab}, \bibinfo{person}{Krzysztof Choromanski},
  \bibinfo{person}{Mario Lucic}, \bibinfo{person}{Yi Tay},
  \bibinfo{person}{Adrian Weller}, {and} \bibinfo{person}{Mostafa Dehghani}.}
  \bibinfo{year}{2021}\natexlab{}.
\newblock \showarticletitle{PolyViT: Co-training Vision Transformers on Images,
  Videos and Audio}.
\newblock \bibinfo{journal}{\emph{arXiv preprint arXiv:2111.12993}}
  (\bibinfo{year}{2021}).
\newblock


\bibitem[Liu et~al\mbox{.}(2020)]%
        {liu2020exploiting}
\bibfield{author}{\bibinfo{person}{Jian Liu}, \bibinfo{person}{Pengpeng Zhao},
  \bibinfo{person}{Fuzhen Zhuang}, \bibinfo{person}{Yanchi Liu},
  \bibinfo{person}{Victor~S Sheng}, \bibinfo{person}{Jiajie Xu},
  \bibinfo{person}{Xiaofang Zhou}, {and} \bibinfo{person}{Hui Xiong}.}
  \bibinfo{year}{2020}\natexlab{}.
\newblock \showarticletitle{Exploiting aesthetic preference in deep cross
  networks for cross-domain recommendation}. In
  \bibinfo{booktitle}{\emph{Proceedings of The Web Conference 2020}}.
  \bibinfo{pages}{2768--2774}.
\newblock


\bibitem[Liu et~al\mbox{.}(2022)]%
        {liu2022collaborative}
\bibfield{author}{\bibinfo{person}{Weiming Liu}, \bibinfo{person}{Xiaolin
  Zheng}, \bibinfo{person}{Mengling Hu}, {and} \bibinfo{person}{Chaochao
  Chen}.} \bibinfo{year}{2022}\natexlab{}.
\newblock \showarticletitle{Collaborative Filtering with Attribution Alignment
  for Review-based Non-overlapped Cross Domain Recommendation}. In
  \bibinfo{booktitle}{\emph{Proceedings of the ACM Web Conference 2022}}.
  \bibinfo{pages}{1181--1190}.
\newblock


\bibitem[Liu et~al\mbox{.}(2021)]%
        {liu2021swin}
\bibfield{author}{\bibinfo{person}{Ze Liu}, \bibinfo{person}{Yutong Lin},
  \bibinfo{person}{Yue Cao}, \bibinfo{person}{Han Hu}, \bibinfo{person}{Yixuan
  Wei}, \bibinfo{person}{Zheng Zhang}, \bibinfo{person}{Stephen Lin}, {and}
  \bibinfo{person}{Baining Guo}.} \bibinfo{year}{2021}\natexlab{}.
\newblock \showarticletitle{Swin transformer: Hierarchical vision transformer
  using shifted windows}. In \bibinfo{booktitle}{\emph{Proceedings of the
  IEEE/CVF International Conference on Computer Vision}}.
  \bibinfo{pages}{10012--10022}.
\newblock


\bibitem[Ma et~al\mbox{.}(2018)]%
        {ma2018modeling}
\bibfield{author}{\bibinfo{person}{Jiaqi Ma}, \bibinfo{person}{Zhe Zhao},
  \bibinfo{person}{Xinyang Yi}, \bibinfo{person}{Jilin Chen},
  \bibinfo{person}{Lichan Hong}, {and} \bibinfo{person}{Ed~H Chi}.}
  \bibinfo{year}{2018}\natexlab{}.
\newblock \showarticletitle{Modeling task relationships in multi-task learning
  with multi-gate mixture-of-experts}. In \bibinfo{booktitle}{\emph{Proceedings
  of the 24th ACM SIGKDD International Conference on Knowledge Discovery \&
  Data Mining}}. \bibinfo{pages}{1930--1939}.
\newblock


\bibitem[Ma et~al\mbox{.}(2019)]%
        {ma2019pi}
\bibfield{author}{\bibinfo{person}{Muyang Ma}, \bibinfo{person}{Pengjie Ren},
  \bibinfo{person}{Yujie Lin}, \bibinfo{person}{Zhumin Chen},
  \bibinfo{person}{Jun Ma}, {and} \bibinfo{person}{Maarten~de Rijke}.}
  \bibinfo{year}{2019}\natexlab{}.
\newblock \showarticletitle{$\pi$-net: A parallel information-sharing network
  for shared-account cross-domain sequential recommendations}. In
  \bibinfo{booktitle}{\emph{Proceedings of the 42nd International ACM SIGIR
  Conference on Research and Development in Information Retrieval}}.
  \bibinfo{pages}{685--694}.
\newblock


\bibitem[Man et~al\mbox{.}(2017)]%
        {man2017cross}
\bibfield{author}{\bibinfo{person}{Tong Man}, \bibinfo{person}{Huawei Shen},
  \bibinfo{person}{Xiaolong Jin}, {and} \bibinfo{person}{Xueqi Cheng}.}
  \bibinfo{year}{2017}\natexlab{}.
\newblock \showarticletitle{Cross-domain recommendation: An embedding and
  mapping approach.}. In \bibinfo{booktitle}{\emph{IJCAI}},
  Vol.~\bibinfo{volume}{17}. \bibinfo{pages}{2464--2470}.
\newblock


\bibitem[Mu et~al\mbox{.}(2022)]%
        {mu2022id}
\bibfield{author}{\bibinfo{person}{Shanlei Mu}, \bibinfo{person}{Yupeng Hou},
  \bibinfo{person}{Wayne~Xin Zhao}, \bibinfo{person}{Yaliang Li}, {and}
  \bibinfo{person}{Bolin Ding}.} \bibinfo{year}{2022}\natexlab{}.
\newblock \showarticletitle{ID-Agnostic User Behavior Pre-training for
  Sequential Recommendation}.
\newblock \bibinfo{journal}{\emph{arXiv preprint arXiv:2206.02323}}
  (\bibinfo{year}{2022}).
\newblock


\bibitem[Ni et~al\mbox{.}(2018)]%
        {ni2018perceive}
\bibfield{author}{\bibinfo{person}{Yabo Ni}, \bibinfo{person}{Dan Ou},
  \bibinfo{person}{Shichen Liu}, \bibinfo{person}{Xiang Li},
  \bibinfo{person}{Wenwu Ou}, \bibinfo{person}{Anxiang Zeng}, {and}
  \bibinfo{person}{Luo Si}.} \bibinfo{year}{2018}\natexlab{}.
\newblock \showarticletitle{Perceive your users in depth: Learning universal
  user representations from multiple e-commerce tasks}. In
  \bibinfo{booktitle}{\emph{Proceedings of the 24th ACM SIGKDD International
  Conference on Knowledge Discovery \& Data Mining}}.
  \bibinfo{pages}{596--605}.
\newblock


\bibitem[Rendle et~al\mbox{.}(2012)]%
        {rendle2012bpr}
\bibfield{author}{\bibinfo{person}{Steffen Rendle}, \bibinfo{person}{Christoph
  Freudenthaler}, \bibinfo{person}{Zeno Gantner}, {and} \bibinfo{person}{Lars
  Schmidt-Thieme}.} \bibinfo{year}{2012}\natexlab{}.
\newblock \showarticletitle{BPR: Bayesian personalized ranking from implicit
  feedback}.
\newblock \bibinfo{journal}{\emph{arXiv preprint arXiv:1205.2618}}
  (\bibinfo{year}{2012}).
\newblock


\bibitem[Sheng et~al\mbox{.}(2021)]%
        {sheng2021one}
\bibfield{author}{\bibinfo{person}{Xiang-Rong Sheng}, \bibinfo{person}{Liqin
  Zhao}, \bibinfo{person}{Guorui Zhou}, \bibinfo{person}{Xinyao Ding},
  \bibinfo{person}{Binding Dai}, \bibinfo{person}{Qiang Luo},
  \bibinfo{person}{Siran Yang}, \bibinfo{person}{Jingshan Lv},
  \bibinfo{person}{Chi Zhang}, \bibinfo{person}{Hongbo Deng}, {et~al\mbox{.}}}
  \bibinfo{year}{2021}\natexlab{}.
\newblock \showarticletitle{One Model to Serve All: Star Topology Adaptive
  Recommender for Multi-Domain CTR Prediction}. In
  \bibinfo{booktitle}{\emph{Proceedings of the 30th ACM International
  Conference on Information \& Knowledge Management}}.
  \bibinfo{pages}{4104--4113}.
\newblock


\bibitem[Shin et~al\mbox{.}(2021b)]%
        {shin2021one4all}
\bibfield{author}{\bibinfo{person}{Kyuyong Shin}, \bibinfo{person}{Hanock
  Kwak}, \bibinfo{person}{Kyung-Min Kim}, \bibinfo{person}{Minkyu Kim},
  \bibinfo{person}{Young-Jin Park}, \bibinfo{person}{Jisu Jeong}, {and}
  \bibinfo{person}{Seungjae Jung}.} \bibinfo{year}{2021}\natexlab{b}.
\newblock \showarticletitle{One4all user representation for recommender systems
  in e-commerce}.
\newblock \bibinfo{journal}{\emph{arXiv preprint arXiv:2106.00573}}
  (\bibinfo{year}{2021}).
\newblock


\bibitem[Shin et~al\mbox{.}(2021a)]%
        {shin2021scaling}
\bibfield{author}{\bibinfo{person}{Kyuyong Shin}, \bibinfo{person}{Hanock
  Kwak}, \bibinfo{person}{Kyung-Min Kim}, \bibinfo{person}{Su~Young Kim}, {and}
  \bibinfo{person}{Max~Nihlen Ramstrom}.} \bibinfo{year}{2021}\natexlab{a}.
\newblock \showarticletitle{Scaling Law for Recommendation Models: Towards
  General-purpose User Representations}.
\newblock \bibinfo{journal}{\emph{arXiv preprint arXiv:2111.11294}}
  (\bibinfo{year}{2021}).
\newblock


\bibitem[Sun et~al\mbox{.}(2019)]%
        {sun2019bert4rec}
\bibfield{author}{\bibinfo{person}{Fei Sun}, \bibinfo{person}{Jun Liu},
  \bibinfo{person}{Jian Wu}, \bibinfo{person}{Changhua Pei},
  \bibinfo{person}{Xiao Lin}, \bibinfo{person}{Wenwu Ou}, {and}
  \bibinfo{person}{Peng Jiang}.} \bibinfo{year}{2019}\natexlab{}.
\newblock \showarticletitle{BERT4Rec: Sequential recommendation with
  bidirectional encoder representations from transformer}. In
  \bibinfo{booktitle}{\emph{Proceedings of the 28th ACM international
  conference on information and knowledge management}}.
  \bibinfo{pages}{1441--1450}.
\newblock


\bibitem[Tang et~al\mbox{.}(2020)]%
        {tang2020progressive}
\bibfield{author}{\bibinfo{person}{Hongyan Tang}, \bibinfo{person}{Junning
  Liu}, \bibinfo{person}{Ming Zhao}, {and} \bibinfo{person}{Xudong Gong}.}
  \bibinfo{year}{2020}\natexlab{}.
\newblock \showarticletitle{Progressive layered extraction (ple): A novel
  multi-task learning (mtl) model for personalized recommendations}. In
  \bibinfo{booktitle}{\emph{Fourteenth ACM Conference on Recommender Systems}}.
  \bibinfo{pages}{269--278}.
\newblock


\bibitem[Van~den Oord et~al\mbox{.}(2018)]%
        {van2018representation}
\bibfield{author}{\bibinfo{person}{Aaron Van~den Oord}, \bibinfo{person}{Yazhe
  Li}, {and} \bibinfo{person}{Oriol Vinyals}.} \bibinfo{year}{2018}\natexlab{}.
\newblock \showarticletitle{Representation learning with contrastive predictive
  coding}.
\newblock \bibinfo{journal}{\emph{arXiv e-prints}} (\bibinfo{year}{2018}),
  \bibinfo{pages}{arXiv--1807}.
\newblock


\bibitem[Wei et~al\mbox{.}(2019)]%
        {wei2019mmgcn}
\bibfield{author}{\bibinfo{person}{Yinwei Wei}, \bibinfo{person}{Xiang Wang},
  \bibinfo{person}{Liqiang Nie}, \bibinfo{person}{Xiangnan He},
  \bibinfo{person}{Richang Hong}, {and} \bibinfo{person}{Tat-Seng Chua}.}
  \bibinfo{year}{2019}\natexlab{}.
\newblock \showarticletitle{MMGCN: Multi-modal graph convolution network for
  personalized recommendation of micro-video}. In
  \bibinfo{booktitle}{\emph{Proceedings of the 27th ACM International
  Conference on Multimedia}}. \bibinfo{pages}{1437--1445}.
\newblock


\bibitem[Wu et~al\mbox{.}(2021a)]%
        {wu2021empowering}
\bibfield{author}{\bibinfo{person}{Chuhan Wu}, \bibinfo{person}{Fangzhao Wu},
  \bibinfo{person}{Tao Qi}, {and} \bibinfo{person}{Yongfeng Huang}.}
  \bibinfo{year}{2021}\natexlab{a}.
\newblock \showarticletitle{Empowering news recommendation with pre-trained
  language models}. In \bibinfo{booktitle}{\emph{Proceedings of the 44th
  International ACM SIGIR Conference on Research and Development in Information
  Retrieval}}. \bibinfo{pages}{1652--1656}.
\newblock


\bibitem[Wu et~al\mbox{.}(2021b)]%
        {wu2021mm}
\bibfield{author}{\bibinfo{person}{Chuhan Wu}, \bibinfo{person}{Fangzhao Wu},
  \bibinfo{person}{Tao Qi}, {and} \bibinfo{person}{Yongfeng Huang}.}
  \bibinfo{year}{2021}\natexlab{b}.
\newblock \showarticletitle{MM-Rec: Multimodal News Recommendation}.
\newblock \bibinfo{journal}{\emph{arXiv preprint arXiv:2104.07407}}
  (\bibinfo{year}{2021}).
\newblock


\bibitem[Wu et~al\mbox{.}(2020)]%
        {wu2020ptum}
\bibfield{author}{\bibinfo{person}{Chuhan Wu}, \bibinfo{person}{Fangzhao Wu},
  \bibinfo{person}{Tao Qi}, \bibinfo{person}{Jianxun Lian},
  \bibinfo{person}{Yongfeng Huang}, {and} \bibinfo{person}{Xing Xie}.}
  \bibinfo{year}{2020}\natexlab{}.
\newblock \showarticletitle{PTUM: Pre-training User Model from Unlabeled User
  Behaviors via Self-supervision}.
\newblock \bibinfo{journal}{\emph{arXiv preprint arXiv:2010.01494}}
  (\bibinfo{year}{2020}).
\newblock


\bibitem[Wu et~al\mbox{.}(2021c)]%
        {wu2021userbert}
\bibfield{author}{\bibinfo{person}{Chuhan Wu}, \bibinfo{person}{Fangzhao Wu},
  \bibinfo{person}{Yang Yu}, \bibinfo{person}{Tao Qi},
  \bibinfo{person}{Yongfeng Huang}, {and} \bibinfo{person}{Xing Xie}.}
  \bibinfo{year}{2021}\natexlab{c}.
\newblock \showarticletitle{UserBERT: Contrastive User Model Pre-training}.
\newblock \bibinfo{journal}{\emph{arXiv preprint arXiv:2109.01274}}
  (\bibinfo{year}{2021}).
\newblock


\bibitem[Yi et~al\mbox{.}(2022)]%
        {yi2022multi}
\bibfield{author}{\bibinfo{person}{Zixuan Yi}, \bibinfo{person}{Xi Wang},
  \bibinfo{person}{Iadh Ounis}, {and} \bibinfo{person}{Craig Macdonald}.}
  \bibinfo{year}{2022}\natexlab{}.
\newblock \showarticletitle{Multi-modal Graph Contrastive Learning for
  Micro-video Recommendation}. In \bibinfo{booktitle}{\emph{Proceedings of the
  45th International ACM SIGIR Conference on Research and Development in
  Information Retrieval}}. \bibinfo{pages}{1807--1811}.
\newblock


\bibitem[Yuan et~al\mbox{.}(2016)]%
        {yuan2016lambdafm}
\bibfield{author}{\bibinfo{person}{Fajie Yuan}, \bibinfo{person}{Guibing Guo},
  \bibinfo{person}{Joemon~M Jose}, \bibinfo{person}{Long Chen},
  \bibinfo{person}{Haitao Yu}, {and} \bibinfo{person}{Weinan Zhang}.}
  \bibinfo{year}{2016}\natexlab{}.
\newblock \showarticletitle{Lambdafm: learning optimal ranking with
  factorization machines using lambda surrogates}. In
  \bibinfo{booktitle}{\emph{Proceedings of the 25th ACM international on
  conference on information and knowledge management}}.
  \bibinfo{pages}{227--236}.
\newblock


\bibitem[Yuan et~al\mbox{.}(2020)]%
        {yuan2020parameter}
\bibfield{author}{\bibinfo{person}{Fajie Yuan}, \bibinfo{person}{Xiangnan He},
  \bibinfo{person}{Alexandros Karatzoglou}, {and} \bibinfo{person}{Liguang
  Zhang}.} \bibinfo{year}{2020}\natexlab{}.
\newblock \showarticletitle{Parameter-efficient transfer from sequential
  behaviors for user modeling and recommendation}. In
  \bibinfo{booktitle}{\emph{Proceedings of the 43rd International ACM SIGIR
  Conference on Research and Development in Information Retrieval}}.
  \bibinfo{pages}{1469--1478}.
\newblock


\bibitem[Yuan et~al\mbox{.}(2019)]%
        {yuan2019simple}
\bibfield{author}{\bibinfo{person}{Fajie Yuan}, \bibinfo{person}{Alexandros
  Karatzoglou}, \bibinfo{person}{Ioannis Arapakis}, \bibinfo{person}{Joemon~M
  Jose}, {and} \bibinfo{person}{Xiangnan He}.} \bibinfo{year}{2019}\natexlab{}.
\newblock \showarticletitle{A simple convolutional generative network for next
  item recommendation}. In \bibinfo{booktitle}{\emph{Proceedings of the Twelfth
  ACM International Conference on Web Search and Data Mining}}.
  \bibinfo{pages}{582--590}.
\newblock


\bibitem[Yuan et~al\mbox{.}(2021)]%
        {yuan2021one}
\bibfield{author}{\bibinfo{person}{Fajie Yuan}, \bibinfo{person}{Guoxiao
  Zhang}, \bibinfo{person}{Alexandros Karatzoglou}, \bibinfo{person}{Joemon
  Jose}, \bibinfo{person}{Beibei Kong}, {and} \bibinfo{person}{Yudong Li}.}
  \bibinfo{year}{2021}\natexlab{}.
\newblock \showarticletitle{One person, one model, one world: Learning
  continual user representation without forgetting}. In
  \bibinfo{booktitle}{\emph{Proceedings of the 44th International ACM SIGIR
  Conference on Research and Development in Information Retrieval}}.
  \bibinfo{pages}{696--705}.
\newblock


\bibitem[Zhao et~al\mbox{.}(2019)]%
        {zhao2019recommending}
\bibfield{author}{\bibinfo{person}{Zhe Zhao}, \bibinfo{person}{Lichan Hong},
  \bibinfo{person}{Li Wei}, \bibinfo{person}{Jilin Chen},
  \bibinfo{person}{Aniruddh Nath}, \bibinfo{person}{Shawn Andrews},
  \bibinfo{person}{Aditee Kumthekar}, \bibinfo{person}{Maheswaran
  Sathiamoorthy}, \bibinfo{person}{Xinyang Yi}, {and} \bibinfo{person}{Ed
  Chi}.} \bibinfo{year}{2019}\natexlab{}.
\newblock \showarticletitle{Recommending what video to watch next: a multitask
  ranking system}. In \bibinfo{booktitle}{\emph{Proceedings of the 13th ACM
  Conference on Recommender Systems}}. \bibinfo{pages}{43--51}.
\newblock


\bibitem[Zhou et~al\mbox{.}(2020)]%
        {zhou2020s3}
\bibfield{author}{\bibinfo{person}{Kun Zhou}, \bibinfo{person}{Hui Wang},
  \bibinfo{person}{Wayne~Xin Zhao}, \bibinfo{person}{Yutao Zhu},
  \bibinfo{person}{Sirui Wang}, \bibinfo{person}{Fuzheng Zhang},
  \bibinfo{person}{Zhongyuan Wang}, {and} \bibinfo{person}{Ji-Rong Wen}.}
  \bibinfo{year}{2020}\natexlab{}.
\newblock \showarticletitle{S3-rec: Self-supervised learning for sequential
  recommendation with mutual information maximization}. In
  \bibinfo{booktitle}{\emph{Proceedings of the 29th ACM International
  Conference on Information \& Knowledge Management}}.
  \bibinfo{pages}{1893--1902}.
\newblock


\bibitem[Zhu et~al\mbox{.}(2020)]%
        {zhu2020deep}
\bibfield{author}{\bibinfo{person}{Feng Zhu}, \bibinfo{person}{Yan Wang},
  \bibinfo{person}{Chaochao Chen}, \bibinfo{person}{Guanfeng Liu},
  \bibinfo{person}{Mehmet Orgun}, {and} \bibinfo{person}{Jia Wu}.}
  \bibinfo{year}{2020}\natexlab{}.
\newblock \showarticletitle{A deep framework for cross-domain and cross-system
  recommendations}.
\newblock \bibinfo{journal}{\emph{arXiv preprint arXiv:2009.06215}}
  (\bibinfo{year}{2020}).
\newblock


\end{thebibliography}
}

\end{document}